\documentclass[journal]{IEEEtran}
\usepackage{algorithm,algorithmic,amsbsy,amsmath,amssymb,epsfig,bbm,mathrsfs,multirow,amsthm}
\usepackage[T1]{fontenc}
\usepackage[latin9]{inputenc}
\usepackage{amsmath}
\usepackage{fixltx2e}
\usepackage{algorithm}
\usepackage{array,multirow,graphicx,epstopdf}
\usepackage{color}
\usepackage{cite}
\usepackage{float}
\usepackage{makecell}
\usepackage[x11names,dvipsnames,table]{xcolor} 
\usepackage{colortbl}
\usepackage{caption,subcaption}
\usepackage{lipsum}
\definecolor{mygray}{gray}{0.6}
\definecolor{myblue}{rgb}{0.8,0.85,1}
\usepackage{array}
\newcolumntype{L}[1]{>{\raggedright\let\newline\\\arraybackslash\hspace{0pt}}m{#1}}
\newcolumntype{C}[1]{>{\centering\let\newline\\\arraybackslash\hspace{0pt}}m{#1}}
\newcolumntype{R}[1]{>{\raggedleft\let\newline\\\arraybackslash\hspace{0pt}}m{#1}}

\usepackage{array, tabularx, boldline}
\usepackage{eurosym}
\usepackage{amstext} 
\DeclareRobustCommand{\officialeuro}{%
  \ifmmode\expandafter\text\fi
  {\fontencoding{U}\fontfamily{eurosym}\selectfont e}}

\usepackage{array, tabularx, boldline}
\usepackage{graphicx}
\usepackage{cellspace}
\begin{document}

\title{\huge Towards Smart Wireless Communications via Intelligent Reflecting Surfaces: A Contemporary Survey}
\author{
Shimin Gong, \textit{Member, IEEE}, Xiao Lu, \textit{Member, IEEE}, Dinh Thai Hoang, \textit{Member, IEEE}, Dusit Niyato, \textit{Fellow, IEEE}, Lei Shu, \textit{Senior Member, IEEE}, Dong In Kim, \textit{Fellow, IEEE},  and Ying-Chang Liang, \textit{Fellow, IEEE}
\thanks{S.~Gong is with School of Intelligent Systems Engineering, Sun Yat-sen University, China. E-mail: gongshm5@mail.sysu.edu.cn.}
\thanks{X.~Lu is with the Department of Electrical and Computer Engineering, University of Alberta, Canada. Email: lu9@ualberta.ca}
\thanks{D.~T.~Hoang is with the Faculty of Engineering and Information Technology,
University of Technology Sydney, Australia. Email: hoang.dinh@uts.edu.au.}
\thanks{D.~Niyato is with School of Computer Science and Engineering, Nanyang Technological University, Singapore. Email: dniyato@ntu.edu.sg.}
\thanks{L.~Shu is with NAU-Lincoln Joint Research Center of Intelligent Engineering, Nanjing Agricultural University, China. Email: lei.shu@njau.edu.cn.}
\thanks{D.~I.~Kim is with the Department of Electrical and Computer Engineering, Sungkyunkwan University, South Korea. Email: dikim@skku.ac.kr.}
\thanks{Y.-C.~Liang is with the Center for Intelligent Networking and Communications (CINC), University of Electronic Science and Technology of China, China. Email: liangyc@ieee.org.}
}


\maketitle
\IEEEpeerreviewmaketitle
\begin{abstract}
This paper presents a literature review on recent applications and design aspects of the intelligent reflecting surface (IRS) in the future wireless networks. Conventionally, the network optimization has been limited to transmission control at two endpoints, i.e., end users and network controller. The fading wireless channel is uncontrollable and becomes one of the main limiting factors for performance improvement. The IRS is composed of a large array of scattering elements, which can be individually configured to generate additional phase shifts to the signal reflections. Hence, it can actively control the signal propagation properties in favor of signal reception, and thus realize the notion of a smart radio environment. As such, the IRS's phase control, combined with the conventional transmission control, can potentially bring performance gain compared to wireless networks without IRS. In this survey, we first introduce basic concepts of the IRS and the realizations of its reconfigurability. Then, we focus on applications of the IRS in wireless communications. We overview different performance metrics and analytical approaches to characterize the performance improvement of IRS-assisted wireless networks. To exploit the performance gain, we discuss the joint optimization of the IRS's phase control and the transceivers' transmission control in different network design problems, e.g.,~rate maximization and power minimization problems. Furthermore, we extend the discussion of IRS-assisted wireless networks to some emerging {use cases}. Finally, we highlight important practical challenges and future research directions for realizing IRS-assisted wireless networks in beyond 5G communications.

\end{abstract}


\section{INTRODUCTION}

With the popularizing of user devices constituting the future Internet of Things (IoT), we have never stopped our efforts on the challenging network optimization problems to improve the energy- or spectrum-efficiency (EE/SE) of wireless networks, with the aim of accommodating the users' demanding data rate and diverse quality of service (QoS) requirements, e.g.,~\cite{tradeoff} and~\cite{eebig18}. Currently, the performance optimization of wireless networks either focuses on the user side or the network controller, e.g., the base station (BS) and network operator. For wireless network operators, the ever-increasing traffic demand can be fulfilled by deploying energy-efficient small cells in a dense network or using multiple antennas at the BS to increase spectrum efficiency~\cite{antenna6g}. The BS's transmit beamforming or power allocation can be optimized to adapt to the channel variations. At the user side, multiple users can join collaboration, e.g.,~via device-to-device (D2D)~\cite{d2d-survey} and relay communications~\cite{multihop}. These features can potentially provide the benefits of improved link quality and coverage, increased EE/SE performance, reduced interference and power consumption~\cite{densenetwork}. A joint optimization can be made possible when the information exchange and coordination between end users and the network controller are available. This is preferred as it generally yields a higher performance gain if it is solvable with affordable cost. Hence, numerous research works in the literature have proposed joint system optimizations to improve the EE/SE performance of wireless networks by a combination of different techniques, e.g.,~\cite{opt_wsn18} and~\cite{opt_pls}. These may include the optimization for wireless power transfer, cooperative relaying, beamforming, and resource allocation, etc.

\begin{figure*}[t]
\centering
\includegraphics[width=0.95\textwidth]{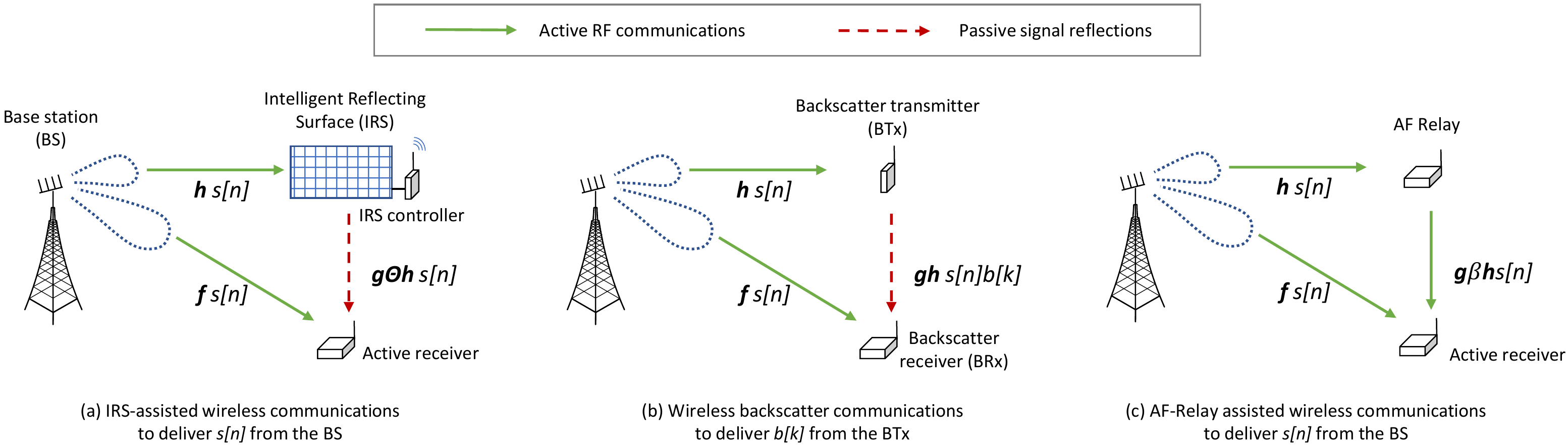}
\caption{Comparing IRS-assisted wireless communications with the backscatter communications and amplify-and-forward (AF) relay-assisted wireless communications. The IRS in (a) introduces a phase shift matrix $\Theta$ to configure the equivalent reflecting channel. The AF relay in (c) introduces a power amplifying coefficient $\beta$ to forward the received signal. The receiver decodes the $n$-th source information symbol $s[n]$ for the IRS-assisted and AF relay-assisted communications, while in (b) it aims at decoding the $k$-th piggybacked information symbol $b[k]$ from the strong interference $s[n]$ in wireless backscatter communications. The BS-IRS, BS-receiver, and IRS-receiver channels are denoted by ${\bf h}$, ${\bf f}$, ${\bf g}$, respectively.}\label{fig:compare}
\end{figure*}

However, in the current paradigm of wireless network optimization, the radio environment itself remains an uncontrollable factor and thus it is not accounted for in the problem formulations. Due to randomness in the radio environment, the signal propagation typically experiences reflections, diffractions, and scattering before reaching the receiver with multiple randomly attenuated and delayed copies of the original signals in different paths. Such a channel fading effect becomes a major limiting factor for the maximization of EE/SE performance of wireless networks. Recently, a novel concept of intelligent reflecting surface (IRS) has been introduced in the wireless communications research community~\cite{18scm_ian3,19survey_renzo,joint_overview}. The IRS is a two-dimensional (2D) man-made surface of electromagnetic (EM) material, namely metasurface, that is composed of a large array of passive scattering elements with specially designed physical structure.\footnote{In the literature, we have observed different names referring to a similar concept of using reconfigurable metasurface to assist wireless communications, e.g.,~software-defined hypersurface, large intelligent surface/antenna, reconfigurable intelligent surface, and holographic MIMO surface. For consistence, we use the term of intelligent reflecting surface (IRS) throughout this paper.} Each scattering element can be controlled in a software-defined manner to change the EM properties (e.g.,~the phase shift) of the reflection of the incident RF signals upon the scattering elements. By a joint phase control of all scattering elements, the reflecting phases and angles of the incident RF signals can be arbitrarily tuned to create a desirable multi-path effect. In particular, the reflected RF signals can be added coherently to improve the received signal power or combined destructively to mitigate interference. A typical system model of IRS-assisted wireless communications is illustrated in Fig.~\ref{fig:compare}(a). By deploying the IRS in the environment, e.g., coated on walls of buildings and carried by aerial platforms, the IRS can turn the radio environment into a smart space that can assist information sensing, analog computing, and wireless communications~\cite{19survey_renzo}. Along with the optimal control of the legacy RF transceivers, IRS-assisted wireless systems will become more flexible to support diverse user requirements, e.g., enhanced data rate, extended coverage, minimized power consumption, and more secure transmissions~\cite{irs_survey,huang2019holographic,liang_survey}.

\begin{table*}
\caption{Using IRS for Realizing the Vision of Smart Radio Environment}
\begin{center}
\begin{tabular}{  l p{25mm} p{45mm} p{30mm} p{45mm}  }
\hline
Reference &  Concept/Scheme & Design approach  & Application scenario  & Validation  \\
\hline
\cite{12smart_wall2} & Intelligent wall  & Switch FSS between ON and OFF to shape the propagation environment & Smart indoor environment for  OFDMA system  & Extend coverage and improve system performance up to 80\% by simulation \\\hline
\cite{12smart_wall1} & ANN-based intelligent wall & Use ANN to explore the optimal setting for controlling the intelligent walls & Smart indoor environment at 2 GHz  & Simulations show quick responses to demands and improved performance\\\hline
\cite{14shaping_wave} & Spatial microwave modulator (SMM) & Use a binary-phase state tunable metasurface to manipulate EM waves  & SMM fabricated at the 2.47 GHz frequency & Experiments show the capability of improving or cancelling RF signals \\\hline
\cite{17smartspace_longfei} & Programmable radio environment &  Embed low-cost devices in walls to passively reflect active RF signals & Indoor 2.4 GHz Wi-Fi-like communications  & Experiments show the efficacy of attenuating or enhancing signal by 26 dB \\\hline
\cite{18scm_ian1} & Programmable wireless environment  & Control current distribution over hypersurface tiles to manipulate EM waves  & Indoor 60 GHz mmWave communications & Simulations demonstrate significantly improved coverage and received power\\\hline
\cite{19pls_ian1} & Programmable wireless environment & Hypersurface for interference control, security, and distortion mitigation & Indoor 2.4 GHz and 60 GHz communications  & Simulations show that groundbreaking performance and security potential \\\hline
\cite{ultra-massive} & Ultra-Massive MIMO & Deploy plasmonic arrays at the transmitter, through the channel, and at the receiver  &  mmWave and THz communications  &  Simulations show significant improvements in communication distance and data rate \\\hline
\cite{optic}  & Intelligent mirror & Create LOS link by rotating the IRS or electronically changing the wavefront & Free space optical (FSO) communications  & Simulations show that building sway for the IRS has either a smaller or larger impact on the channel quality\\\hline
\cite{19ch_diversity}  & Physical shaping of propagation medium &  Install binary-phase state tunable metasurface inside random environment  &  2.47 GHz Wi-Fi frequency  & Experiments show perfect channel orthogonality, optimal channel diversity, flexible interference suppression \\\hline
\end{tabular}
 \end{center} \label{tab-smartwall}
 \end{table*}

The vision of {\em smart radio environment} has been demonstrated via realistic simulations and experiments to verify its capability of improving transmission performance in different wireless networks leveraging the IRS's reconfigurability. The {\em active wall} is introduced in~\cite{12smart_wall2,12smart_wall1} by using an active frequency-selective surface (FSS) to manipulate the wireless environment. The FSS provides a narrow-band frequency filtering of incoming signals~\cite{Chang2008Equivalent}, which can be used to build up a cognitive engine that makes the walls intelligent. The experiments in~\cite{17smartspace_longfei} demonstrate that, by deploying hybrid active-passive elements in the walls of a building, we can expect to 1) mitigate the destructive effect of multiple paths for legacy wireless communications, 2) eliminate poorly conditioned similar channels in a large-scale MIMO system, and 3) simultaneously attenuate interference power and increase signal strength at different receivers. The idea of {\em programmable wireless channels or environment} is proposed in~\cite{18scm_ian1,19pls_ian1} by using hypersurfaces, i.e., software-controlled metamaterials, to cover physical objects in the radio environment. Several physical layer building-block technologies for a programmable wireless environment are implemented and evaluated in~\cite{19modeling_ian}. By controlling the distribution of the current over the hypersurfaces, the incident RF signals can be reshaped with a desirable response, thus realizing a reconfigurable wireless environment. By alleviating signal path loss, multi-path fading, and co-channel interference, the programmable wireless environment is capable of improving transmission performance in terms of signal quality, communication range, and EE/SE performance. Different from the use of hypersurfaces, the authors in~\cite{ultra-massive} introduce Ultra-Massive MIMO platforms to realize an intelligent communication environment, which enables full-wave control with plasmonic arrays deployed at the transmitter, through the channel, and at the receiver. Although massive MIMO can provide remarkable improvement to the wireless channel conditions, the EM manipulation of massive MIMO is only feasible at the transceivers, while the wireless environment remains passive and uncontrollable. As a result, the channel model is still a probabilistic process, rather than a software-defined reconfigurable counterpart that is actively participating in the communication process. A summary of different concepts and realizations for a smart radio environment is listed in Table~\ref{tab-smartwall}. The application scenarios are rather diverse, from media sharing Wi-Fi systems to mmWave, THz, and even optical communications, covering a broadband frequency range, while the common observation is that the IRS can be utilized to create desirable radiation patterns in favor of wireless communications.

By integrating the smart radio environment into the network optimization problems, IRS-assisted wireless networks are envisioned to revolutionize the current network optimization paradigm and expected to play an active role in future wireless networks~\cite{antenna6g,19learning_renzo}. The specific benefits of IRS-assisted wireless communications can be summarized as follows:
\begin{itemize}
  \item {\bf Easy deployment and sustainable operations:} The IRS is made of low-cost passive scattering elements embedded in the metasurface. It can be in any shape, thus providing high flexibility for its deployment and replacement. It can be easily attached to and removed from facades of buildings, indoor walls, and ceilings, etc. Without the use of active components for power-consuming signal processing algorithms, the IRS can be battery-less and wirelessly powered by RF-based energy harvesting.
 \item {\bf Flexible reconfiguration via passive beamforming:} The IRS can bring additional phase shifts to the reflected signals. By jointly optimizing the phase shifts of all scattering elements, namely passive beamforming, the signal reflections can be coherently focused at the intended receiver and nulled at other directions. The number of reflecting elements can be extremely large, e.g.,~from tens to hundreds~\cite{irs_survey}, depending on the traverse size of the IRS. This implies a great potential for performance improvement of wireless networks. The IRS's phase control, combined with the transceivers' operational parameters, e.g.,~transmit beamforming, power allocation, and resource allocation, can be jointly optimized to explore the performance gain of IRS-assisted wireless networks.
  \item {\bf Enhanced capacity and EE/SE performance:} By using the IRS, the wireless channel can be programmed to support a higher link capacity with reduced power consumption for point-to-point communications. Interference suppression also becomes effective by using the IRS, which implies a better signal quality for the cell-edge users. For multi-user (MU) wireless networks, the scattering elements can be partitioned and allocated to assist data transmissions of different users. As such, the IRS-assisted wireless network can provide better QoS provisioning and potentially improve the sum-rate performance or max-min fairness among different users.
  \item {\bf Exploration of emerging wireless applications:} The development of the IRS is expected to pave the way for new promising research directions. For example, the IRS has recently been introduced to be a novel approach for preventing wireless eavesdropping attacks by simultaneously controlling the transmission at the transmitter and the reflections at the IRS. Many other emerging research areas also benefit from the use of the IRS such as wireless power transfer, unmanned aerial vehicle (UAV) communications, and mobile edge computing (MEC), which will be reviewed in this survey.
\end{itemize}

Although the IRS's operation resembles that of a multi-antenna relay, it is fundamentally different from the existing relay communications. By using passive elements, the IRS can realize fully controllable beam steering without dedicated energy supply and sophisticated active circuitry for channel estimation, information decoding, and amplifying and forwarding. Compared to the conventional amplify-and-forward (AF) relay that actively generates new RF signals, the IRS does not use an active transmitter but only reflects the ambient RF signals as a passive array, which incurs no additional power consumption. It also differs from the conventional backscatter communications, where the backscatter transmitter communicates with its receiver by modulating and reflecting the ambient RF signals~\cite{backscatter18}. The backscattered information is piggybacked in the ambient RF signals. Despite their differences, the IRS can also be used to perform wireless backscatter communications due to its capability of manipulating the phases of the reflected signals, e.g.,~\cite{kim,liangback,channel-backscatter}. The evolving path from backscatter radios to large intelligent surfaces is revealed in~\cite{liang_survey}. The difference among three communication technologies is illustrated in Fig.~\ref{fig:compare}. Such a different feature of IRS-assisted wireless networks thus motivates the necessity of presenting a comprehensive literature review, which aims at providing fundamental knowledge about the IRS's physical properties and implementations, diverse applications in different network scenarios, and the performance optimizations of IRS-assisted wireless networks.

\begin{figure}[t]
\centering
\includegraphics[width=0.48\textwidth]{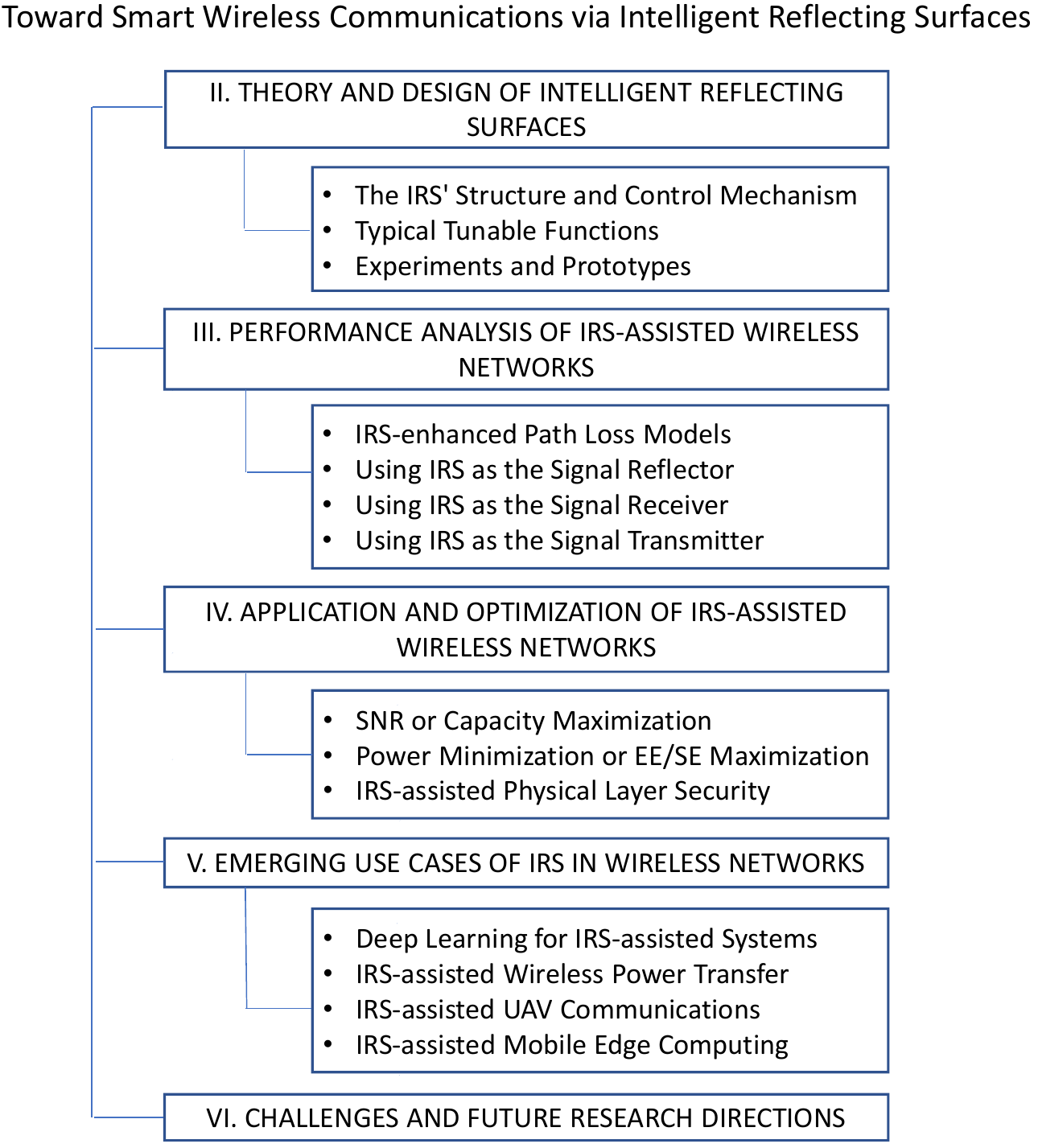}
\caption{Survey framework and outline of the main topics.}\label{fig:outline}
\end{figure}

There are a few review papers in the literature, either focusing on the physical design and implementations of the reconfigurable metasurface, e.g.,~\cite{12surface_review,metasurface16,17surface_review}, or its conceptual applications in wireless communications and networking, e.g.,~\cite{19survey_renzo,irs_survey,huang2019holographic}. In particular, the review papers~\cite{12surface_review,metasurface16,17surface_review} mainly focus on the theoretical basis, physics characterization, and classification of metasurfaces as well as their applications at different operational frequencies. Motivated by the IRS's appealing EM properties, the authors in~\cite{19survey_renzo} discuss the feasibility and methodologies of using the IRS to realize the concept of smart radio environments, and also analyze potential solutions to some fundamental challenges towards its massive deployment and applications in future wireless networks. Following on, the review paper in~\cite{irs_survey} provides a more technical overview with a special focus on the analysis of signal models and the physical layer channel enhancement in IRS-assisted wireless communications. The authors in~\cite{huang2019holographic} introduce the concept of holographic multiple-input multiple-output (MIMO) surface (HMIMOS) comprising sub-wavelength metallic or dielectric scattering particles, which shares a similar idea with the use of the IRS in wireless communications. An overview of HMIMOS communications is provided by introducing the hardware architectures, classifications, and main characteristics. Several short survey and tutorial papers also appear in the literature, e.g.,~\cite{joint_overview,liang_survey}, mainly discussing the recent applications of the IRS in typical wireless scenarios. The tutorial in~\cite{joint_overview} covers an overview of the IRS technology to create smart and reconfigurable environment, including its main applications in wireless communication, competitive advantages over existing technologies, hardware architecture as well as the corresponding new signal model, with the focus on the key challenges in designing and implementing the IRS-assisted wireless networks. Liang \emph{et al.} in~\cite{liang_survey} firstly gave a tutorial on the evolving from the backscatter communications, reflective relay, to intelligent reflective surfaces, which are all relying on the use of inexpensive and passive components to realize enhanced wireless information transmission by using EM scattering principles. The tutorial covers a comparison of the signal models and design aspects of different radio technologies. Comparing to these works, our survey offers a comprehensive study on the theoretical basis of the IRS and a contemporary review on its most recent applications in wireless networks. Specifically, our major contributions are summarized as follows:
\begin{itemize}
  \item A systematic organization of the literature under extensive review is provided (as shown in Fig.~\ref{fig:outline}) for readers with different backgrounds and interests to comprehend the IRS technology more effectively. We start from the physical characterization of the IRS and its EM properties, covering both design methodologies and experimental prototypes. Then we focus on IRS-assisted wireless networks by classifying different papers according to their design objectives and control variables.
  \item We provide an extensive review on the stochastic analysis of performance limit and asymptotic behavior of IRS-assisted wireless networks, which are not covered by the existing review papers. The pervasive deployment of the IRS will change the random nature of the channel environment, which calls for different analytical tools and performance metrics to characterize the performance limits. Corroborated by the stochastic analysis, the potential increase in performance gain then motivates the further optimization of IRS-assisted wireless networks.
  \item Some common shortcomings of the current literature are analyzed and technical insights are highlighted for future exploration. For example, the current optimization frameworks for IRS-assisted wireless networks mainly rely on the alternating optimization method, which implies that a higher performance gain can be achieved potentially with a more sophisticated algorithm design. Besides, a majority of the papers in the literature focus on joint active and passive beamforming optimization. In fact, the passive beamforming can also be jointly optimized with other control strategies, e.g.,~information encoding, transmission scheduling, access control, and full-duplex communications. We also notice that the energy consumption of the IRS is usually omitted in the literature, which may lead to over-optimistic conclusions.
\end{itemize}

\begin{table}[h!]
\scriptsize
  \caption{\small List of abbreviations}
  \label{tab:abbr}\centering
  \begin{tabularx}{8.7cm}{|Sl|X|}
    \hline
  \cellcolor{mygray} \textbf{Abbreviation} &   \cellcolor{mygray} \textbf{Description} \\ \hline
3D/2D & 3-Dimensional/2-Dimensional \\\hline
ADMM & Alternating Direction Method of Multipliers \\\hline
AF/DF & Amplify/Decode-and-Forward \\\hline
AI & Artificial Intelligence\\\hline
ANN/DNN &Artificial/Deep Neural Network\\ \hline
BCD & Block Coordinate Descent\\    \hline
BER/SER & Bit Error Rate/Symbol Error Rate\\ \hline
BS/AP &Base Station/Access Point \\\hline
CSI &Channel State Information\\\hline
CRN & Cognitive Radio Network\\\hline
D2D &Device-to-Device\\ \hline
DC & Difference-of-Convex \\\hline
DL/RL & Deep/Reinforcement Learning \\\hline
EE/SE & Energy- or Spectrum-Efficiency\\\hline
FPGA  & Field-Programmable Gate Array \\\hline
FSS & Frequency-Selective Surface\\ \hline
GHz/THz & Gigahertz/Terahertz \\\hline
HetNets & Heterogeneous Networks \\\hline
IC/NoC &  Integrated Circuit/Network-on-Chip\\\hline
IoT & Internet of Things\\\hline
IRS & Intelligent Reflecting Surface \\\hline
LOS & Line-of-Sight \\ \hline
LT/LR & Legitimate Transmitter/Legitimate Receiver \\\hline
MAC & Medium Access Control\\\hline
MEC & Mobile Edge Computing \\\hline
MISO/MIMO & Multiple-Input and Single-Output/Multiple-Output  \\ \hline
MM& Majorization Minimization\\   \hline
mmWave & Millimeter Wave \\\hline
MU & Multi-user \\ \hline
OFDM &Orthogonal Frequency Division Multiplexing\\\hline
OFDMA & Orthogonal Frequency Division Multiple Access \\\hline
QoS& Quality of Service\\ \hline
RB & Resource Block \\ \hline
RF/EM & Radio Frequency/Electromagnetic\\\hline
SCA & Successive Convex Approximation \\\hline
SDR &Semidefinite Relaxation \\\hline
SNR/SINR &Signal-to-Noise Ratio/Signal-to-Interference plus Noise Ratio\\\hline
SM/SMM &Spatial Modulation/Spatial Microwave Modulator\\\hline
SOCP & Second-Order Cone Programming \\\hline
SWIPT & Simultaneous Wireless Information and Power Transfer \\\hline
TDMA/NOMA & Time Division/Non-Orthogonal Multiple Access \\\hline
UAV& Unmanned Aerial Vehicle\\\hline
ZF & Zero-Forcing \\\hline
\end{tabularx}
\end{table}
The rest of this paper is organized as follows. Section~\ref{sec_intro} presents the basic theory and implementation of the IRS. After this, this paper focuses on IRS-assisted wireless networks. Section~\ref{sec_analysis} discusses the path loss and signal models, performance metrics, and provides a stochastic analysis of IRS-assisted wireless networks, considering the random deployment of scattering elements in the radio environment. Section~\ref{sec_optimization} presents the performance optimization of IRS-assisted wireless networks, typically by a joint optimization of the IRS's passive beamforming and the transceivers' transmit control. Section~\ref{sec_misc} discusses more emerging use cases of the IRS in wireless communications. Finally, we summarize some practical challenges and future research directions in Section~\ref{sec_open}, and then conclude the paper in Section~\ref{sec_cons}. The main topics covered in this paper are shown in Fig.~\ref{fig:outline}. A list of abbreviations used in this survey is given in Table~\ref{tab:abbr}.

\section{THEORY AND DESIGN OF INTELLIGENT REFLECTING SURFACES} \label{sec_intro}

The metasurface is a kind of two-dimensional (i.e., with near-zero thickness) man-made material that exhibits special EM properties depending on its structural parameters. As illustrated in Fig.~\ref{fig:metasurface}, the metasurface is composed of a large array of passive scattering elements, e.g.,~metallic or dielectric particles, that can transform the impinging EM waves in different ways~\cite{dielectric}. The sub-wavelength structural arrangement of the scattering elements determines how the incident waves are transformed, i.e., the direction and strength of the reflected and diffracted waves. In general, when EM waves propagate to a boundary between two different medias, the strengths and directions of the reflected and diffracted waves typically follow the Fresnel equations and Snell's law, respectively~\cite{metasurface16}. The situation becomes different when the same wave impinges upon a metasurface. The periodical arrangement of the scattering elements can cause a shift of the resonance frequency and thus a change of boundary conditions. As a result, the reflected and diffracted waves will carry additional phase changes.

Once the metasurface is fabricated with a specific physical structure, it will have fixed EM properties and therefore can be used for a specific purpose, e.g.,~a perfect absorber operating at a certain frequency. The analysis of EM properties can be based on the general-purpose full-wave EM simulator or approximate computational techniques~\cite{approx_design18}. More efficient analytical approaches rely on sophisticated boundary conditions to describe the metasurface discontinuity and its EM responses~\cite{18surface_analysis,18surface_nature}. However, it becomes very inflexible as a new metasurface has to be re-designed and fabricated to serve another purpose or operate at a different frequency. In particular, based on the application requirements, the structural parameters of the scattering elements constituting the metasurface have to be recalculated by a synthesis approach~\cite{15surface_tensor,18surface_prop3}, which is in general computational demanding.

The IRS is built from a reconfigurable metasurface, which can fully control the phase shifts incurred by individual scattering elements. This can be achieved by imposing external stimuli on the scattering elements and thus alter their physical parameters, leading to the change of EM properties of the metasurface without refabrication~\cite{tuning18}. The first design issue for the IRS lies in a control mechanism to connect and communicate with a large size of scattering elements, and thus agilely and jointly control their EM behaviors on demand. The other main issue is the realization of reconfigurability to achieve full and accurate phase control of the reflected or diffracted waves. In the sequel, we first discuss the IRS's structural design and inter-cell communications mechanism to connect and control all scattering elements. After that, we discuss and compare different phase tuning mechanisms to achieve a variety of tunable functions and their applications. We also provide a review on prototypes and experiments in the literature to verify the feasibility of using the IRS in practice.

\subsection{The IRS's Structure and Control Mechanism}

\subsubsection{IRS Controller and Tunable Chips}

\begin{figure}[t]
\centering
\includegraphics[width=0.48\textwidth]{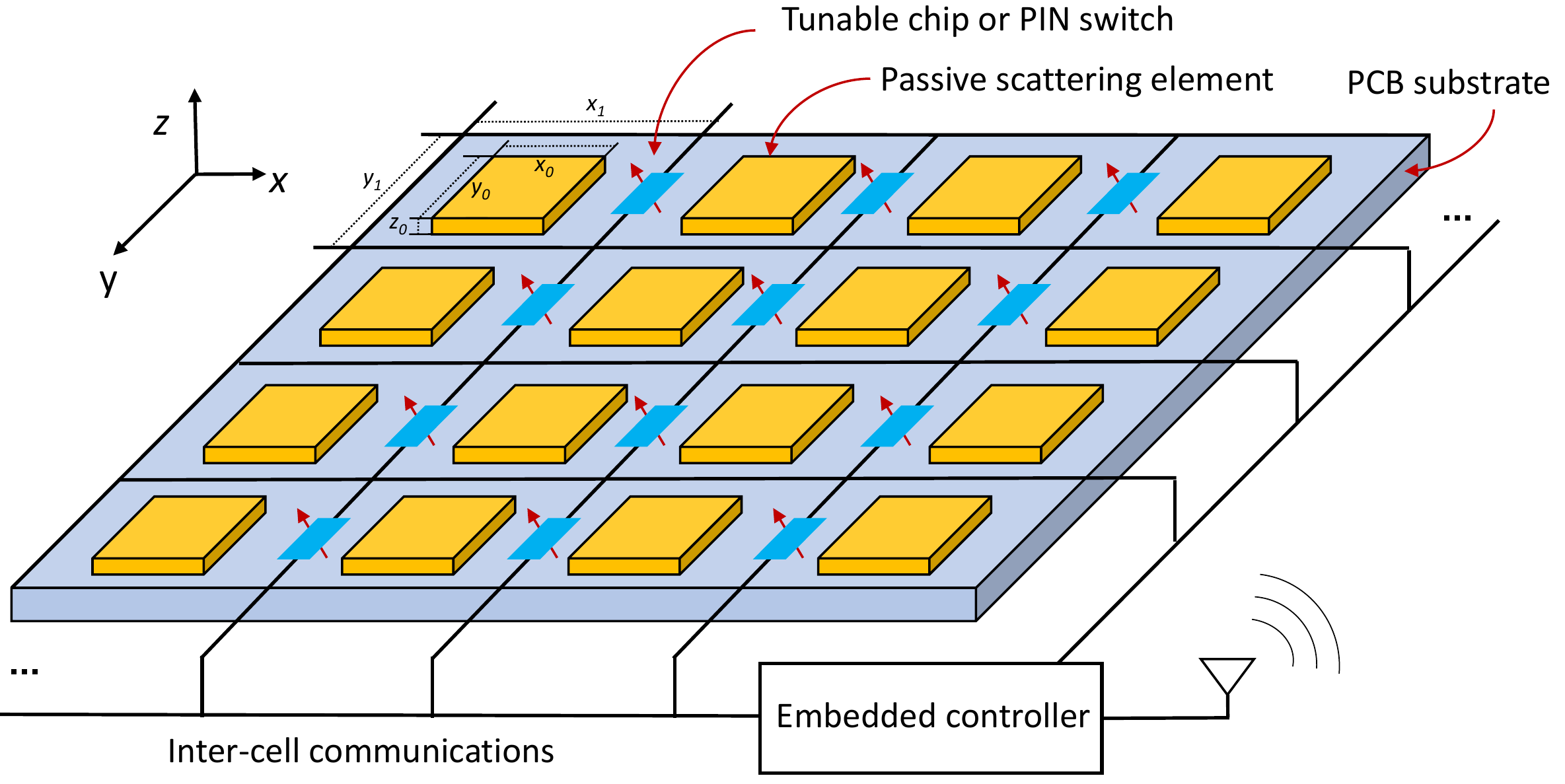}
\caption{The IRS is made of a reconfigurable metasurface composed of a large array of passive scattering elements.}\label{fig:metasurface}
\end{figure}

In general, the IRS's reconfiguration of its EM behaviors is achieved by a joint phase control of individual scattering elements. This implies an integration of tunable chips within the structure of the metasurface, where each tunable chip interacts locally with a scattering element and communicates to a central controller, e.g.,~\cite{16surface_focusing,tuning18}. Hence, it allows a software-defined implementation of the control mechanism~\cite{nano15}. For example, the IRS controller can be implemented in a field-programmable gate array (FPGA) and the tunable chips are typical PIN diodes~\cite{16surface_focusing}. As illustrated in Fig.~\ref{fig:metasurface}, the embedded IRS controller can communicate and receive reconfiguration request from external devices, and then optimize and distribute its phase control decisions to all tunable chips. Upon receiving the control information, each tunable chip changes its state and allows the corresponding scattering element to reconfigure its behavior. The IRS can be also equipped with embedded sensors with the capability of sensing the environment~\cite{19survey_renzo}. Such sensing information can be used by the IRS controller to automatically update its configuration and thus maintain consistent EM behaviors under dynamic environmental conditions.

The tunable chips can be PIN diodes with ON and OFF states. This allows the change of input impedance to match or mismatch with the free space impedance~\cite{tuning18}. The authors in~\cite{14surface_design} design and demonstrate through experiments a binary-state tunable chip based on the hybridized resonator controlled by a PIN diode as the unit cell of the metasurface, operating with the resonance frequency around 2.466 GHz. The tunable chips can also be varactor diodes, which can be adjusted in a continuous way given different voltage bias, e.g.,~\cite{varacitor13,kim16}. The integrated circuits (ICs) with continuously tunable load impedance are designed in~\cite{19surface_review1} to control the phase shifts of scattering elements. Both tunable resistance and capacitance in the ICs can be controlled by imposing upon a gate voltage. Hence, wave manipulation can be realized by optimizing the biasing voltages. As an example, the authors in~\cite{tuning18} demonstrate the perfect absorption at 5 GHz for different polarizations and incidence angles.

The authors in~\cite{nano15} and~\cite{17surface_survey} propose to integrate a network of tiny controllers within the metasurface and wirelessly connect it to an external device. Each controller is capable of interpreting external instructions and tuning its varactor to achieve a desired impedance configuration. The change of connectivity at different locations of the controller network can realize the reconfiguration of the IRS's physical structure, resulting in multiple tunable functionalities. In contrast, the authors in~\cite{18surface_survey2} design the IRS by connecting the scattering elements to a smaller group of controller chips. Each controller chip serves four metallic scattering elements. The controller chips can adjust the EM behaviors of the metasurface by attributing additional local resistance and reactance on demand. Similar to~\cite{18surface_survey2}, the metasurface designed in~\cite{18surface_verification} is composed of reconfigurable metamaterial strips arranged in a grid. A set of four strips is controlled by an intra-tile controller (i.e., tunable chip). All intra-tile controllers are interconnected to constitute the intra-tile network, which can receive external configuration instructions from gateway controllers.

\subsubsection{Inter-cell Communications}
The IRS's reconfigurability depends on the inter-cell communications among the tunable chips jointly controlling the scattering elements of the metasurface to exhibit desirable tunable functions. Communications among the underlying chip controllers can be either wired or wireless~\cite{18surface_survey2}. Wired communication can be preferable as it can be integrated with the controllers within the same chip. Wireless inter-cell communication becomes a compelling alternative in either large-scale or dense metasurfaces. The design of inter-communications protocols has to comply with stringent energy, latency, and robustness requirements~\cite{18surface_routing}.

The authors in~\cite{18surface_survey2} propose two approaches for wireless inter-cell communications. The first approach exploits the metasurface structure, while the second approach employs a dedicated communication channel beneath the metasurface structure. In the first approach, the communication channel is the space between the scattering elements and the substrate, which acts as a waveguide for signal propagation. The second approach is achieved by adding an extra metallic plate below the chip. The dedicated communication channel provides an obstacle-free waveguide for information communication between tunable chips. To ensure reliable communications among controllers embedded in the metasurface structure, the authors in~\cite{18surface_routing} adopt the traditional Network-on-Chip (NoC) methodologies and develop two fault adaptive routing algorithms, which can bypass the faulty links by using a properly designed fault-tolerant routing metric.

The authors in~\cite{18surface_circuit} develop embedded ICs within the metasurface structure, which uses mini-routers to move control information among an arbitrarily-large size of scattering elements. All ICs are connected to a shared gateway, which can receive configuration instructions from external wireless devices. Each IC can configure the load impedance at different locations of the metasurface. The authors design a handshaking mechanism to coordinate the information exchange among different metasurface cells. Such a design can achieve salient benefits, including delay insensitivity, low EM emissions, and low power consumption. Medium access control (MAC) strategies are investigated in~\cite{18mac_chip} to share the wireless medium efficiently among metasurface cells. The analysis of physical constraints, performance objectives, and traffic characteristics of on-chip communications shreds some insights on the MAC protocol designs for a large number of metasurface cells.

\subsubsection{Phase Tuning Mechanism}
The IRS's reconfigurability depends on manipulating the phase of individual scattering elements. When the external or ambient stimulus changes, the physical parameters of the scattering elements and substrate will be tuned accordingly. Typical stimulus includes electric, magnetic, light, and thermal stimulus, which can tune the main body of a metasurface and thus provide a global control over its EM properties, e.g., absorption level, resonance frequency, and polarization of waves. Individual phase control of each scattering element is also possible, by applying the stimuli locally to each scattering element. This method is expected to achieve more sophisticated wave manipulations, such as beam steering, focusing, imaging, and holography~\cite{tuning18}.

\begin{figure}[t]
\centering
\includegraphics[width=0.48\textwidth]{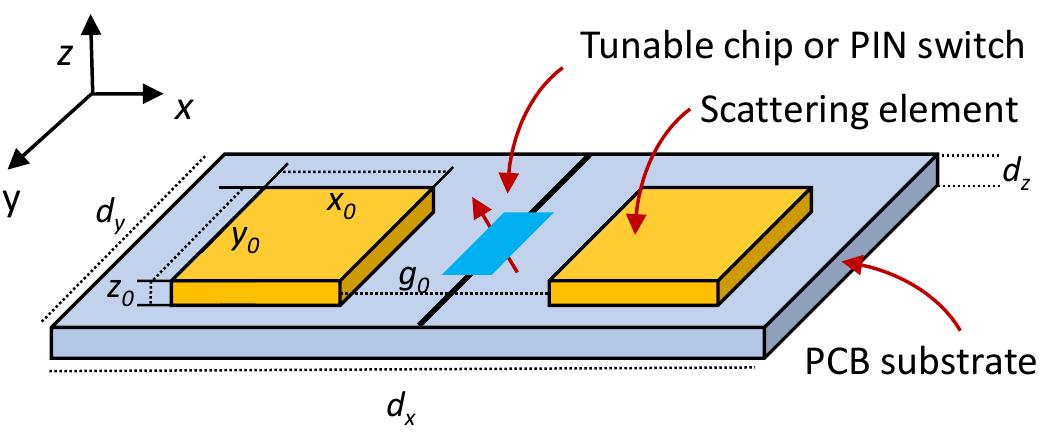}
\caption{Schematic of the unit cell for IRS, e.g.,~\cite{19surface_review1}. The basic topology and dimension parameters $(d_x, d_y, d_z, x_0, y_0, z_0, g_0)$ can be selected for operations in different frequency ranges. A tunable chip (e.g.,~diode and varactor) is incorporated to provide a variable impedance in a continuous way.}\label{fig:unitcell}
\end{figure}

The most straightforward way to achieve local tuning is by changing the physical dimension of the scattering element, resulting in the change of resonant frequency and hence the phase shift, as illustrated in Fig.~\ref{fig:unitcell}. The authors in~\cite{15surface_tera} implement the full controllable phase shifts by combining two different scattering elements. The overall wave reflections can be minimized by optimizing the array pattern. More prevailing tuning approaches are based on the electrically controlled binary-phase tuning or continuous reactance tuning mechanism, e.g., by using diodes or varactors, respectively, e.g.,~\cite{varacitor13,kim16,19surface_review1}. The authors in~\cite{17surface_liquid} use the electronically controllable liquid crystal for real-time wave manipulation. Each metasurface cell is loaded with a thin layer of the liquid crystal. By controlling the voltage bias on each cell, the effective dielectric constant will change and therefore lead to desirable phase shifts at various locations of the metasurface. Full-wave simulation results verify that effective beam steering can be achieved in real time. The authors in~\cite{18surface_prop2} experimentally verify that the use of an acoustic cell architecture can provide enough degrees of freedom to control a refractive metasurface. In particular, a normal incident wave can be redirected over $60^{\circ}$ with the efficiency up to 90\%.

\subsection{Typical Tunable Functions}

The IRS can support a wide range of tunable functions, such as perfect absorption, anomalous reflection, beam shaping, and steering~\cite{17surface_review,18surface_prop1,16surface_focusing}, as illustrated in Fig.~\ref{fig:tunable}. Moreover, it is capable of sensing and communicating with the external devices~\cite{19survey_renzo}. This allows it to be integrated with wireless communication systems and utilized in a variety of wireless applications, e.g.,~\cite{irs_survey,joint_overview,wireless_application}.

\subsubsection{Perfect Absorption} The phase shifts of a metasurface can be designed to ensure minimal reflected and/or refracted power of the impinging waves. Such a metasurface can be used to design an ultra-thin invisibility skin cloak, typically used for visible light~\cite{18surface_nature,cloaking18}. In particular, absorbing metasurface can be used as carpet cloaking of a scatterer on a flat ground plane~\cite{cloaking15}. This is achieved by ensuring the same reflection angle as the incident angle everywhere on the metasurface. The authors in~\cite{15surface_tera} propose a single-layer terahertz (THz) metasurface to produce ultra-low reflections across a broad-frequency spectrum and wide incidence angles. The authors in~\cite{kim16} show that perfect absorption can be achieved at an operating frequency in the range of 4-6 GHz by applying a change in the reverse biasing voltage of the varactor. In~\cite{17surface_review}, the authors introduce the design of broadband absorbers of light by using unit cells featuring with multiple adjacent resonances. A tunable metasurface absorber is presented in~\cite{18surface_optical} by using an optically-programmable capacitor as the tunable chip of each cell. The designed metasurface can operate at 5.5 GHz and achieve a tuning bandwidth of 150 MHz. The authors in~\cite{19surface_review1} realize a reconfigurable metasurface that exhibits perfect absorption at 5 GHz with different incidence angles, by changing the capacitance of each cell.

\subsubsection{Anomalous Reflection}
Abnormal reflection can be observed when light beams impinge upon optical metasurfaces~\cite{light11,13surface_control,optic_meta17}. The authors in~\cite{light11} realize anomalous reflection of light beams in optically thin arrays of metallic antennas on silicon. The authors in~\cite{13surface_control} study the optical properties of different metasurfaces and present a metasurface design that can reflect two orthogonal polarization states in a broad wavelength range. The authors in~\cite{optic_meta17} design optical metasurfaces to facilitate tuning of the reflection phase and polarization. The experiment demonstrates the feasibility of tuning the reflection phase over $\pi$ by controlling the resonant properties of the antennas. An acoustic metasurface is constructed in~\cite{13surface_acoustic}, which can tailor the reflected waves with discrete phase shifts covering the full $2\pi$ span. The authors in~\cite{17surface_control} show that perfect phase control can be realized by completely eliminating the metasurface's parasitic reflections into unwanted directions. The resulting power efficiency of reflection can be over 90\% at 8 GHz. A lossless metasurface is proposed in~\cite{18surface_bmcontrol}, where the incident wave can be perfectly received and converted into a surface wave along the surface, before it is radiated into space without power loss from a different location from the receiving point.

\begin{figure}[t]
\centering
\includegraphics[width=0.45\textwidth]{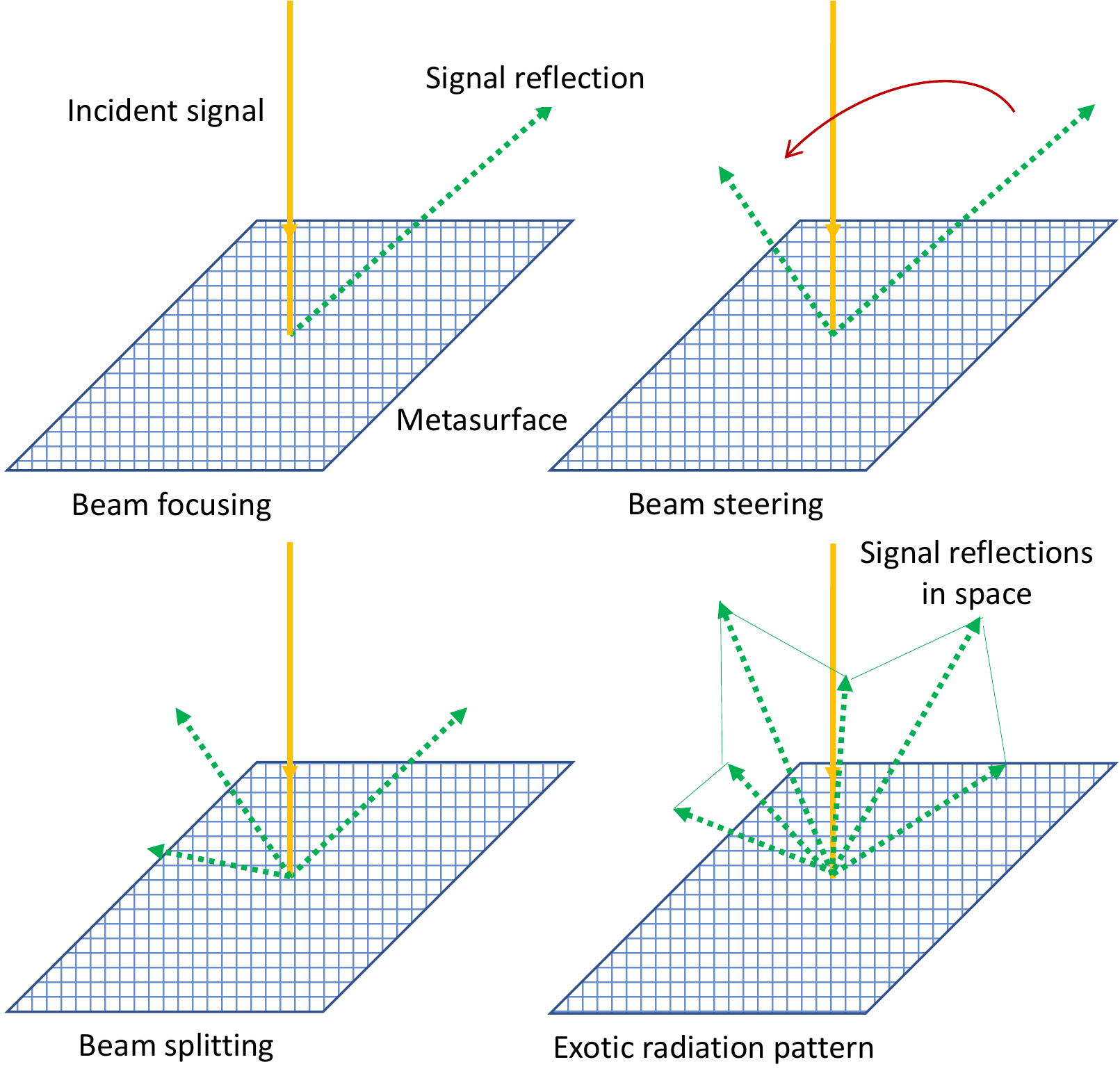}
\caption{Illustrations of typical tunable functions.}\label{fig:tunable}
\end{figure}

\subsubsection{Wave Manipulation} Wave manipulation/modulation can create multiple reflections in different directions simultaneously based on perfect phase control of the metasurface. That is, the power of the reflected waves can be temporal-spatially distributed to create an exotic power radiation pattern, as shown in Fig.~\ref{fig:tunable}, which can be exploited to carry information by spatial modulation~\cite{indexmod16}. Using PIN diode as binary-state control of each scattering element, the authors in~\cite{16surface_focusing} optimize the binary coding matrix and thus create different wave manipulations of a large metasurface, including anomalous reflection, diffusion, beam steering and beamforming. The real-time switch among these radiation patterns can be achieved by an FPGA-based controller. A genetic algorithm is employed in~\cite{17surface_modulating} for arbitrary wave modulations to create desirable radiation patterns according to application requirements. The effectiveness of the genetic algorithm is experimentally verified at an operating frequency of 10 GHz, showing that the accuracy of wave modulation increases with the size of scattering elements. Different from the aforementioned spatial wave modulations, the authors in~\cite{18coding_nature} verify the possibility of simultaneous wave manipulations in both space and frequency domains. The perfect control of the reflection angles and power distribution can be achieved by using a space-time modulated digital coding metasurface.

\subsubsection{Analog Computing} It is also possible for the IRS to perform more complicated mathematical operations (such as spatial differentiation, integration, convolution, and even neural network training) as the impinging wave propagates through the scattering elements. This is referred to as wave-based analog computing, achieving a higher energy efficiency compared to conventional digital signal processing paradigms. The authors in~\cite{14science} introduce the concept of metamaterial analog computing that uses optical metasurfaces to perform mathematical operations in the spatial Fourier domain. This offers the possibility of miniaturized, potentially integrable, wave-based computing systems. Analog computing of acoustic metasurfaces is also demonstrated in~\cite{acoustic_math18} by using thin planar metamaterials to perform mathematical operation in spatial domain. This is promising for various applications including high throughput image processing, ultra-fast equation solving, and real-time signal processing. The authors in~\cite{18analog_comput} experimentally demonstrate that the off-the-shelf wireless infrastructure in combination with a tunable binary-phase metasurface can perform analog computation with Wi-Fi signals.

Some other exotic tunable functions also appear in the literature. The authors in~\cite{sensing11} use the metasurface directly as advanced sensing devices for diagnostic applications, e.g., cancer detection, biological tissue characterization and chemical analysis, based on the interactions of EM waves with the metasurface. The authors in~\cite{storage13} show the possibility of storage and retrieval of EM waves by introducing varactor diodes to manipulate the metasurface's structure. The authors in~\cite{storage17} demonstrate that the slowdown, storage and retrieval of multi-mode EM waves can be achieved through active manipulation of a control field, which shows the possibility for multi-mode memory of EM waves and its practical applications in information processing.

\begin{table*}[t]
\caption{Summary of Experiments and Prototypes of Reconfigurable Metasurfaces}
\begin{center}
\begin{tabular}{  l p{20mm} p{20mm} p{25mm} p{20mm}  p{60mm} }
\hline
Reference & Controller & Dimension & Phase Control & Frequency range & Realized Tunable Functions \\\hline
\cite{16surface_focusing} & FPGA  &  $40\times 40$ array    & Binary PIN diodes & 9-12 GHz & Wave manipulation including anomalous reflection, diffusion, shaped scattering  \\\hline
\cite{18surface_prop2}  & --  &  7 cells & -- & 3.0 GHz & Redirecting a normal incident wave to $60^{\circ}$, $70^{\circ}$, and $80^{\circ}$, with efficiency over 90\% \\\hline
\cite{wireless_application}  & FPGA & $16\times 16$ & 2-bit IRS element & 2.3 GHz, 28.5 GHz & 21.7 dBi antenna gain obtained at 2.3 GHz, 19.1 dBi antenna gain achieved at 28.5 GHz mmWave \\\hline
\cite{13surface_acoustic} &  -- &  8 cells & Traverse length & Acoustic & Discrete phase shifts covering the full $2\pi$ span with steps of $\pi/4$  \\\hline
\cite{17surface_modulating} &  -- &  $40\times40$ array & Cell layout &  8.7-11.3 GHz & Generating different radiation patterns. Even distribution of reflection phases from $0$ to $2\pi$. \\\hline
\cite{18coding_nature} & FPGA & $8\times8$ array & Binary PIN diodes & 8-10 GHz & Simultaneous wave manipulations in both space and frequency domains \\\hline
\cite{irsptype17} & DC voltage source &  $22\times22$ array  & PIN switches and varactor diodes  & 11.5-13.5 GHz & Dynamical beam deflection, splitting, and polarization. 180$^{\circ}$ reflection phase difference\\\hline
\cite{thz_metamaterial19} &  -- & $900\times900$ cm  & Geometric parameters & 0.6-0.9 THz & Broadband reflector with a bandwidth of 0.15 THz and efficiency up to 95\%  \\ \hline
\cite{xbandant12} &   --  &  244 cells & Binary PIN diodes & 10.10-10.70 GHz & Beam switching between $-5^{\circ}$ and $5^{\circ}$ \\
\hline
\cite{ra12} &  --  &  $6\times6$ array & Varactor diodes & 5 GHz & Beam
scanning over a $100^{\circ}\times100^{\circ}$ window \\\hline
\cite{absorber10} & Bias network  & $21\times 21$ cm  & Binary PIN diodes & 2.75-4.0 GHz& Switching between total reflection and absorption  \\\hline
\cite{18smart_mmwave} & MEGA 2560 micro-controller & $14 \times 16$ array &  Binary relay switches & 60 GHz (IEEE 802.11ad)  & Robust link between transceivers can be established by steering the incident signal to the desired receiver \\\hline
\end{tabular}
\end{center} \label{tab:prototypes}
\end{table*}

\subsection{Experiments and Prototypes}

Prototypes of reconfigurable metasurfaces have been developed to verify the feasibility of different tunable functions. Table~\ref{tab:prototypes} summarizes recent experiments and prototypes developed in the literature for the verification. The authors in~\cite{16surface_focusing} implement a metasurface containing 1600 individually controllable cells to demonstrate the feasibility of dynamic wave manipulations. Each cell of the metasurface is integrated with one PIN diode that can switch between two states. The authors in~\cite{18surface_prop2} propose and experimentally verify the use of an acoustic cell architecture to provide enough degrees of freedom for full phase control. Three refractive metasurfaces are designed to redirect a normally incident plane wave by $60^{\circ}$, $70^{\circ}$, and $80^{\circ}$, respectively, with the efficiency up to 90\%. The authors in~\cite{16surface_expment} design a graphene-integrated metasurface to induce a tunable phase change to the incident wave, which can be controlled at an ultra-fast speed. The designed prototype is shown to change the reflection phase up to 55 degrees. The authors in~\cite{thz_metamaterial19} develop a large-scale THz all-dielectric metamaterials with the outer dimension 900 cm $\times$ 900 cm by using the template-assisted fabrication method. Using such metamaterials, the authors implement a broadband reflector with a bandwidth of 0.15 THz and demonstrate its reflection up to 95\%, which implies a wide variety of applications in low-loss and high-efficiency THz devices.

The development of reconfigurable metasurfaces shares a similar idea with the classical concept of reconfigurable reflectarrays, in which the resulting radiation pattern of the signal transmitter is altered as desired by changing the current distribution~\cite{irs_survey}. Comparing to reconfigurable reflectarrays, IRS is featured with the real-time control and reconfigurability of the radio environment~\cite{antenna6g}. A reconfigurable reflectarray antenna is implemented in~\cite{xbandant12}, consisting of a group of 244 radiating elements phase controlled by PIN diodes. The antenna is designed to operate in the band from 10.10 GHz to 10.70 GHz, capable of switching the beam between $-5^{\circ}$ and $5^{\circ}$. Different from the PIN diodes in~\cite{xbandant12}, the authors in~\cite{ra12} implement a $6 \times 6$ transmitarray controlled by varactor diodes to verify its beamforming capability. Experiments demonstrate its capability of beam scanning over a $100 \times 100$-degree window at the operating frequency of 5 GHz. To reduce the size of transmitarray, the authors in~\cite{compact17} design and verify a compact reconfigurable antenna for wireless communications, which is capable of generating four different radiation patterns at the operating frequency of 2.45 GHz.

VISORSURF is an interdisciplinary program funded by Horizon 2020 FET-OPEN~\cite{18surface_verification,h2020}. Its objective is to develop a hardware platform for reconfigurable metasurfaces, namely hypersurface, featured with programmable EM behavior by controlling a network of switches. A prototype of the reconfigurable metasurface is implemented in~\cite{irsptype17} to achieve multi-functional control of EM waves, e.g., beam splitting, deflection, and abnormal reflection at microwave frequencies. This is achieved by controlling the PIN switches and varactor diodes associated with each scattering element. The real-time switch between different EM functionalities is controlled by a computer-controlled voltage source. Realizing wave manipulations in both space and frequency domains, the authors in~\cite{18coding_nature} design a space-time modulated digital coding metasurface to control the propagation direction and power distribution simultaneously in the frequency range of 8-10 GHz. The experimental results demonstrate a good performance for beam steering, beam shaping, and scattering-signature control. The authors in~\cite{wireless_application} develop an IRS prototype with 256 reflecting elements and verify its performance gains in wireless communications, i.e., a 21.7 dBi antenna gain can be obtained at 2.3 GHz and a 19.1 dBi antenna gain can be achieved at 28.5 GHz mmWave frequency. The power consumption by using the IRS can also be reduced significantly. Reference~\cite{18smart_mmwave} implements a 60 GHz reconfigurable reflect-arrays to assist IEEE 802.11ad-based mmWave communications, creating additional reflection links when there exist no line-of-sight (LOS) links.

In the following, we focus on the applications of the IRS in wireless communications. We first introduce the recent works discussing the novel concept of a smart/programmable wireless environment, which is envisioned to change the wireless networking paradigm by using the IRS in wireless communications. As such, we review the system modeling, performance analysis, and optimizations of IRS-assisted wireless networks.

\section{PERFORMANCE ANALYSIS OF IRS-ASSISTED WIRELESS NETWORKS}\label{sec_analysis}

The vision of the smart radio environment can be realized by coating IRS to the facades of physical objects in the radio environment, such as the walls and ceilings, or even carrying IRS in aerial platforms, such as floating balloons and UAVs~\cite{aerial}. This implies that the IRS's scattering elements are distributed in nature relating to the spatial distribution of the environmental objects. Therefore, the modeling and performance analysis of IRS-assisted wireless networks necessitate analytical models that take into account i) the spatial locations of IRS units, ii) the IRS's EM properties, and iii) the wave manipulations adopted by the coexisting IRS units in the environment. In this part, we first review two typical channel models for IRS-assisted wireless communications, and then present the analytical studies on performance limit of IRS-assisted wireless networks, including performance bounds and asymptotic behaviors that are hard to obtain from simulations. Such a theoretical performance analysis can provide design insights on the deployment and configuration of IRS-assisted wireless networks without the need for extensive simulations.

\subsection{IRS-enhanced Path Loss Models}

\begin{figure}[t]
\centering
\includegraphics[width=0.5\textwidth]{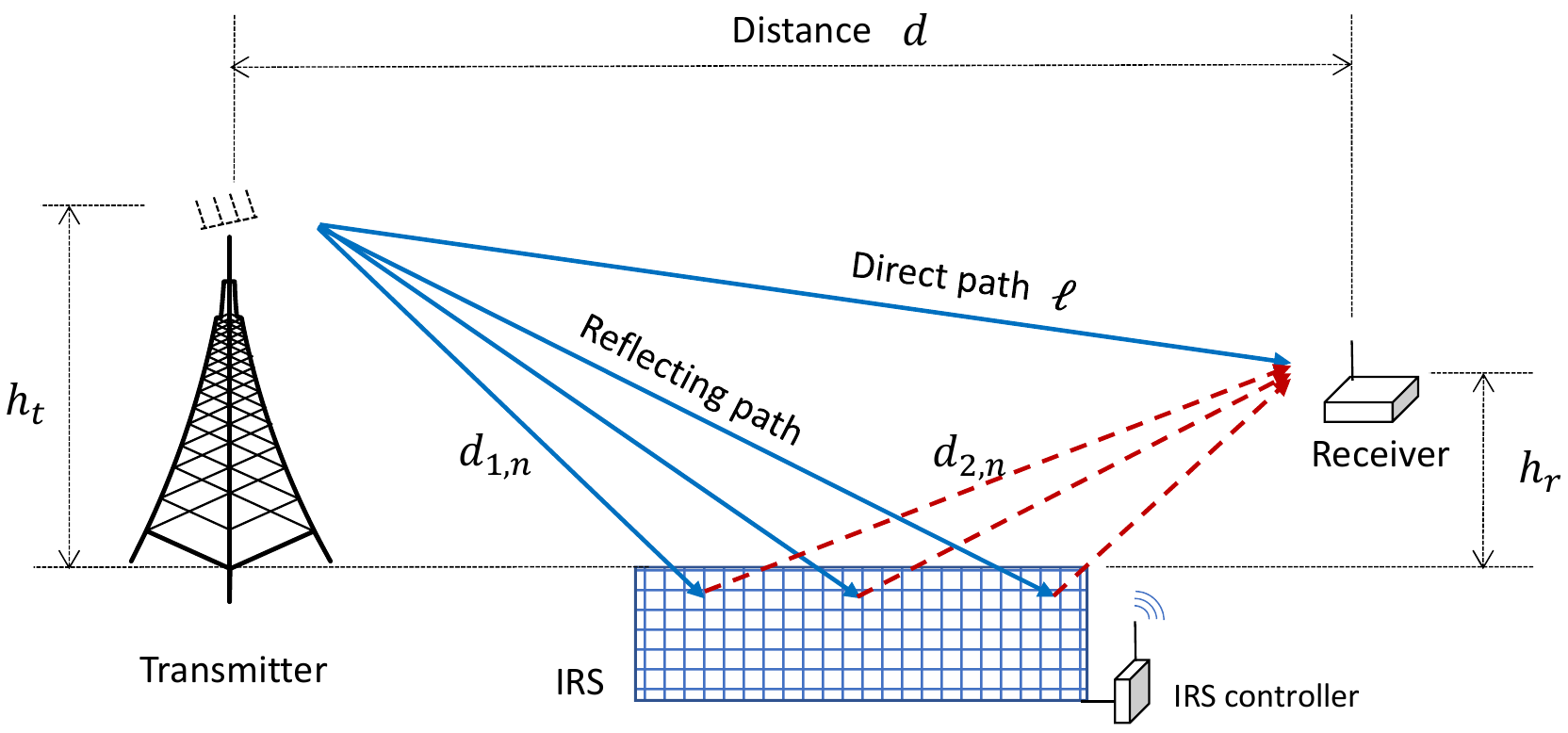}
\caption{Two-way channel model for IRS-assisted wireless communications.}\label{fig_channel}
\end{figure}

The potential performance gain of using IRS in wireless networks can be verified by developing a path loss model to analyze the received signal power and SNR performance. The authors in~\cite{irs_survey} derive a simplified path loss model based on the conventional two-way channel model for wireless communications. As shown in Fig.~\ref{fig_channel}, in addition to the direct path, each IRS's reflecting element provides a second path from the transmitter to the receiver, constituting a two-way signal propagation model. Combining the signals from all paths, the received signal power can be formulated as follows:
\begin{equation}\label{equ-power}
P_r = P_t \left(\frac{\lambda}{4\pi}\right)^2\left|\frac{1}{\ell}+\sum_{n=1}^N\frac{R_n e^{-j\Delta \phi_n}}{d_{1,n}+d_{2,n}}\right|^2,
\end{equation}
where $N$ denotes the number of IRS's reflecting elements and $\ell$ denotes the distance of direct path, which can be approximated by the distance $d$ between the transceivers. $d_{1,n}$ and $d_{2,n}$ denote the distances of two segments in the reflected path via the $n$-th reflecting element. The summation term in~\eqref{equ-power} denotes the signal reflections via different paths. The phase difference $\Delta \phi_n$ is determined by the distances of the direct path and the reflected path via the $n$-th reflecting element. $R_n$ denotes the reflection coefficient depending on the EM properties of the reflecting object, which is conventionally uncontrollable without using IRS.

For each reflecting element on the IRS, we can proactively control its phase shift $R_n = e^{j\Delta \phi_n}$ such that the reflected signal can be coherently aligned with the direct path. Normally, we can assume $d\approx\ell \approx d_{1,n}+d_{2,n}$, which leads to the following approximation of the received signal power in~\eqref{equ-power}.
\begin{equation}\label{equ-power-app}
P_r \propto (N+1)^2 P_t \left(\frac{\lambda}{4\pi d}\right)^2.
\end{equation}
If there is no direct link or the size $N$ of reflecting elements becomes large, the above path loss can be simply rewritten as $P_r \approx N^2 P_t \left(\frac{\lambda}{4\pi d}\right)^2$. Compared to the free-space path loss model, the IRS-assisted path loss in~\eqref{equ-power-app} introduces additional gain to the received signal power, i.e.,~the received power is proportional to $N^2$. For an IRS with 100 reflecting elements, the power gain amounts to a significant 40 dB. The authors in~\cite{irs_survey} also derive the SNR performance in a single-antenna point-to-point system under flat fading channel conditions. For a large number of scattering elements with optimal phase configuration, the compound channel from the transmitter to the receiver can be viewed as a Gaussian distributed random variable with the mean and variance proportional to $N$. This implies that the average received SNR will increases proportionally to $N^2$. The distribution of SNR can also be characterized and used to derive the symbol error probability of IRS-assisted wireless communications. The upper bound of the symbol error probability is shown to have a waterfall region when SNR is low while fall into a slowly-decaying region for higher SNR. In both regions, the error probabilities can be decreased significantly by using a larger IRS.

The simplified path loss model in~\eqref{equ-power-app} fosters basic understanding of IRS's superiority in wireless communications, however omits complex physical details during the IRS-assisted signal propagation, e.g., the size and dimension of the reflecting elements, the angles of incident and reflected signals, and the anisotropic radiation pattern of antennas at the transceivers. Based on a detailed study on the IRS's physics and EM nature, the authors in~\cite{tang2019wireless} propose a more reliable free-space path loss model for IRS-assisted wireless communications. Extensive simulation results validate that the proposed channel model matches well with the experimental measurement results conducted in a microwave anechoic chamber. In particular, given the transmit power and the positions of transceivers, the optimal received signal power in the far-field can be characterized as follows:
\begin{equation}\label{equ-phy-channel}
P_r \propto G_c N^2 P_t \left(\frac{\lambda}{4\pi d_1 d_2}\right)^2  \left(\frac{\rho^2 F(\theta_t,\phi_t)F(\theta_r,\phi_r) }{4\pi}\right),
\end{equation}
where $d_1$, $d_2$ denote the distance from the transmitter to the center point of the IRS, and from the center point to the receiver, respectively. The function $F(\cdot,\cdot)$ determines the antennas' power radiation pattern at the transceivers. In particular, $F(\theta_t,\phi_t)$ returns the normalized power level when the transmitter has the elevation and azimuth angles given by $(\theta_t,\phi_t)$, with respect to the IRS's center point. The constant $\rho$ denotes that all reflecting elements use the same magnitude of reflection coefficients. From the analytical results in~\cite{tang2019wireless}, the coefficient $G_c$ depends on the antenna gains at the transceivers, the dimension and power radiation pattern of the unit reflecting element. Similar to~\eqref{equ-power-app}, the path loss model in~\eqref{equ-phy-channel} implies that in both cases the received signal power is proportional to $N^2$. A different observation is that the far-field received signal power in~\eqref{equ-phy-channel} depends on the product-distance, i.e.,~it is proportional to~$1/(d_1d_2)^2$, instead of the sum-distance $1/(d_1+d_2)^2$ in~\eqref{equ-power-app}. However, the analytical results in~\cite{tang2019wireless} reveal that the sum-distance rule can hold for the received signal power in the radiative near-field. A similar result has been revealed in~\cite{pathloss}.

The signal model of IRS-assisted wireless communications can be established based on the two-way channel model. Considering a simple case with single antenna at the transceivers, let $f_{1,n}$ and $f_{2,n}$ denote the complex channels before and after the reflection point, $g$ be the complex channel for the direct link, the IRS-assisted channel from the transmitter to the receiver is thus given by
\[
\tilde{g} = g  + \sum_{n=1}^N \rho_n \exp{(j\phi_n)} f_{2,n}f_{1,n} = g + {\bf f}_{1}^H \Theta {\bf f}_{2},
\]
where $\rho_n$ and $\phi_n$ represent the magnitude and complex phase of the reflection coefficient at each reflecting element, which can be tuned properly such that the reflected signal is aligned with the direct channel $g$. ${\bf f}_1$ and ${\bf f}_2$ denote the channel vectors of $f_{1,n}$ and $f_{2,n}$. The matrix $\Theta$ has the diagonal vector given by $[\rho_1 \exp{(j\phi_1)},\rho_2 \exp{(j\phi_2)},\ldots,\rho_N \exp{(j\phi_N)}]^H$, which denotes the reflecting coefficients of each reflecting element. For a multi-antenna case, the IRS-assisted channel can be characterized in a similar form. Most of the existing analytical works are built based on the above path loss and signal models, which are basis for the following performance analysis and optimal design of IRS-assisted wireless systems.

\subsection{Using IRS as the Signal Reflector}

The performance metrics of interest mainly include the probabilistic metrics to characterize the uncertainty in wireless transmissions of individual transceivers, as well as the ergodic metrics to characterize the averaged network performance, considering the randomness in network topology, the distribution of reflectors, channel conditions, and interference variations, etc. Typical performance metrics proposed for IRS-assisted wireless systems include the following aspects.
\begin{itemize}
  \item {\em Reflection probability}: The probability that an IRS can reflect the signals from a transmitter to the receiver.
  \item {\em Coverage or outage probability}: The probability that the received SNR is above or below a target threshold.
  \item {\em Bit error probability}: The probability that the decoded information bit differs from the transmitted one.
  \item {\em Ergodic capacity}: The expectation of channel capacity measured by Shannon's formula.
  \item {\em Transport capacity}: The aggregated data rate that can be reliably communicated in the entire system.
\end{itemize}

\subsubsection{Probabilistic Performance}

The {\em reflection probability} is studied in~\cite{19ref_prob_renzo} considering a large-scale IRS randomly distributed by a Boolean model of line segments~\cite{14blockage}. The authors derive the exact probability that a random located IRS can reflect for a given transceiver according to the generalized laws of reflection. The analytical results reveal that the reflection probability of a randomly located reflector is independent of its length. However, this work assumes that all the IRSs have a fixed length, which cannot capture the real-world network environment. Besides, the authors only analyze the reflection probability, without an evaluation on the improvement of transmission performance by using the large-scale IRS. The authors in~\cite{18smart_mmwave} propose the use of the IRS to improve the {\em LOS probability} for indoor mmWave communications. Compared to the existing relay-based approaches that create LOS links for mmWave communications, the reflection-based approach exempts from self-interference of a full-duplex relay and compromised throughput of a half-duplex relay.

Considering the downlink transmission from an mmWave-based access point (AP) to a user with only the reflection links (i.e., no direct link due to obstacles), the authors in~\cite{18smart_mmwave} derive an expression for the {\em outage probability} and further minimize the outage probability by optimizing the IRS's deployment position. The authors in~\cite{outage_multi_irs} study the outage probability of IRS-assisted systems under Rician fading where the IRS's phase shifts only adapt to the LOS components. The outage performance is firstly analyzed and optimized in the slow fading scenario for the non-LOS components. It can be shown that the optimal outage probability decreases with the size of the IRS when the LOS components are stronger than the non-LOS ones. Then, the authors characterize the asymptotically optimal outage probability in the high SNR regime, and show that it decreases with the powers of the LOS components.

The {\em coverage probability} is studied for an IRS-assisted wireless network in~\cite{17coverage}. Different from~\cite{18smart_mmwave} which studies point-to-point mmWave communications, the authors in~\cite{17coverage} consider a generalized mmWave downlink cellular network coexisting with random obstacles and reflectors. A stochastic geometry method is proposed to analyze the downlink coverage probability under the assumption that the locations of base stations follow a homogeneous Poisson point process, the blockages and reflectors are deployed in straight line segments with uniformly distributed orientation and length. The study indicates that only the deployment of reflectors with high density can cause a noticeable improvement in the mmWave coverage. However, the deployment with low density may not benefit mmWave coverage as the reflected signals go through longer distances than the direct links. A limitation of this work is that only the reflections from the nearest reflector is considered. In fact, reflections from other nearby reflectors could also be strong and should not be ignored.


In contrast to above efforts that focus on an analytical study, reference~\cite{18errorpb_renzo1} presents a joint analytical and empirical study for the {\em bit error probability} of spatial modulation based on reconfigurable antennas, which encodes information bits on the radiation patterns of a reconfigurable antenna. To evaluate the impact of the radiation patterns on the error performance, the authors in~\cite{18errorpb_renzo1} introduce an analytical framework to characterize the bit error probability and identify the best radiation pattern to minimize the average error performance. In~\cite{ber-noma}, the BER performance is derived in closed form for an IRS-assisted non-orthogonal multiple access (NOMA) downlink system. The authors in~\cite{symbol-error} present a general mathematical framework to derive the symbol error probability of an IRS-assisted wireless system, where the IRS acts either as a reflector or an access point with a simple transceiver architecture. The numerical results verify that ultra-reliable communications can be achieved by using the IRS to boost the received SNR.

\begin{figure}
\centering
\includegraphics[width=0.4\textwidth]{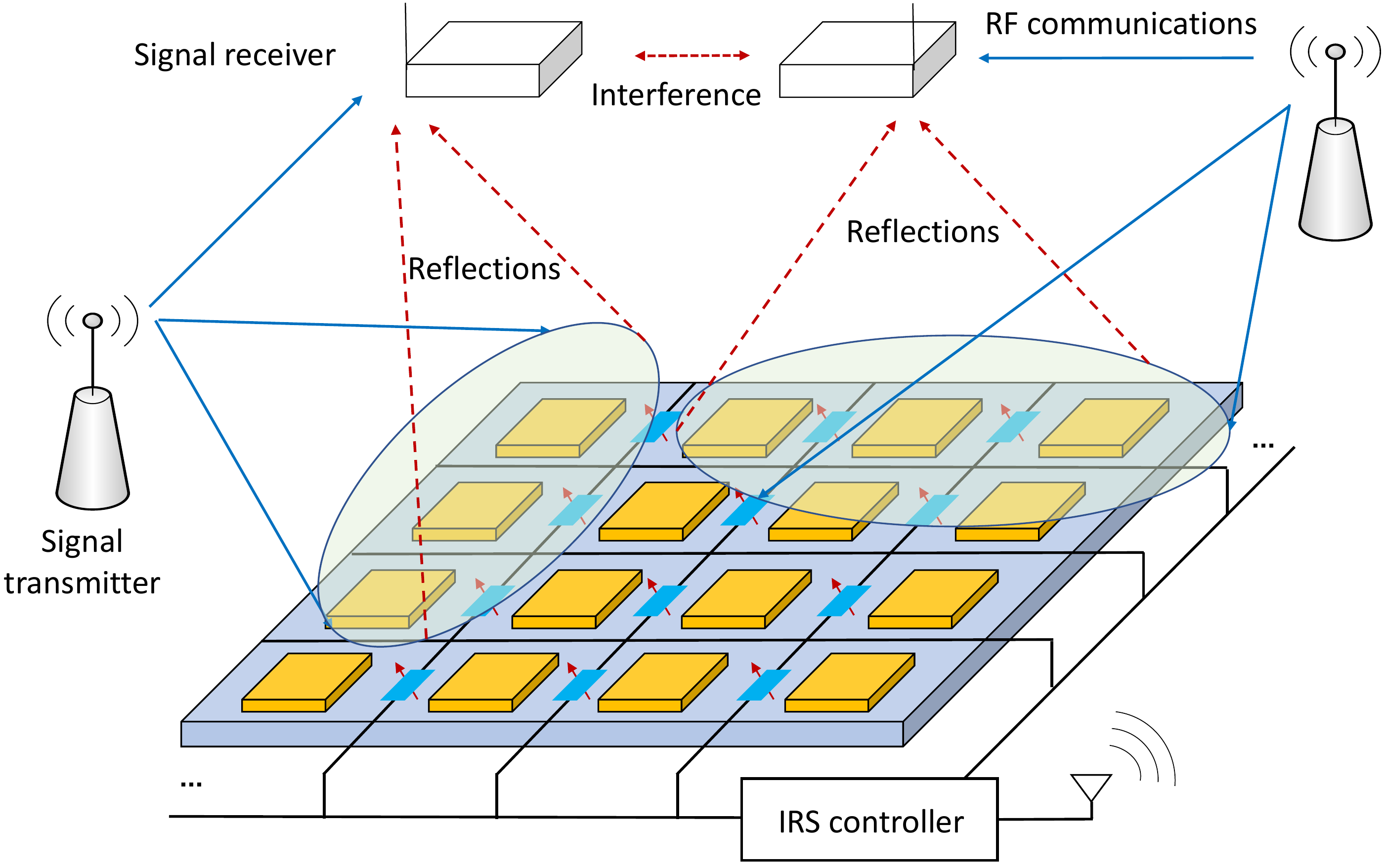}
\caption{The IRS is used for enhancing spectrum sharing among multiple transceivers.}\label{fig_sharing}
\end{figure}

\subsubsection{Ergodic Capacity and Data Rate}

The authors in~\cite{optimality} analyze the {\em asymptotic achievable rate} in an IRS-assisted downlink system. A passive beamformer is designed to achieve the optimal asymptotic performance by tuning the EM properties of signal waves. To maximize the data rate, a modulation scheme is designed for the IRS that is interference free for existing devices. The authors also derive the expected asymptotic symbol error rate (SER) of the considered system and propose a protocol for joint user scheduling and power control. Simulation results verify that the achievable rate in a practical IRS-assisted system satisfies the asymptotic optimality. Considering an IRS-assisted mmWave MIMO system, the authors in~\cite{he2019adaptive} characterize the achievable data rate from the BS and to a mobile user, by designing the IRS's optimal phase shifts based on limited feedback from the mobile user. Besides improvement on data rate, simulation results show that the positioning error bound and orientation error bound both can be reduced by using the IRS with perfect CSI.

From another aspect, the authors in~\cite{rank-improve} consider achieving capacity gain of a point-to-point MIMO system by optimizing the IRS's phase matrix to improve the rank of channel matrix. The IRS-assisted rank improvement enriches the propagation conditions by adding more multi-paths with different spatial angles. With 100 reflecting elements, the numerical results show that the capacity gain can be over 100\% by carefully deploying the IRS's location and optimizing its phase matrix. The aforementioned works focus on the use of ideal IRS with infinite phase resolution, and thus the resulting capacity analysis bears an undetermined gap with the practice. The authors in~\cite{phaseshifts} derive an approximation of the achievable data rate and characterize the performance degradation when a practical IRS is implemented with limited phase shifts. Considering hardware impairments in the IRS's reflecting elements, the authors in~\cite{correlated} model the correlation structure of the hardware impairments by a function of the distance between reflecting elements and analyze the degradation of achievable rate by using a simple receiver structure.

Different from the prior works that focus on link-level capacity maximization for IRS-aided wireless systems, the authors in~\cite{spatial} characterize the spatial throughput of a single-cell MU downlink system assisted by multiple randomly distributed IRSs. The spatial throughput is averaged over the distributions of all users and IRSs' random locations. The simulation results show that the IRS-assisted system can achieve higher spatial throughput than that of a full-duplex relay system. To maximize the spatial throughput, the authors reveal an interesting throughput-fairness tradeoff, which shows that it is preferable to have fewer IRSs with more reflecting elements, however at the cost of more diverse distribution of user rates. The authors in~\cite{capacity-region} analyze the capacity region of a multiple access channel where two users are assisted by IRS units to send information to an AP. The deployment of IRS units can be either distributed or centralized. The distributed IRS units are closer to the users while the centralized IRS is in the vicinity of the AP. For both deployment schemes, capacity regions can be characterized and compared analytically. It can be shown that the centralized scheme outperforms the distributed scheme under the practical channel setup. Both works verify that the IRS-assisted system generally achieves higher capacity performance, which becomes more significant with asymmetric user rates. Hence, we may envision that the IRS has the superiority for the fulfillment of diverse user requirements in future wireless networks.

For multiple transceiver pairs coexisting in one system, the capacity analysis becomes more involved as it depends on the multiple access protocols. The authors in~\cite{16sharing_capacity} propose the IRS-assisted spectrum sharing scheme for multiple transmitters. As shown in Fig.~\ref{fig_sharing}, the idea is to use the IRS for steering the signal beams from different transmitters to enhance the useful signals and cancel the interference towards their respective receivers. This paradigm allows multiple transmitters to simultaneously operate on the same spectrum band without causing interference to each other. Considering an indoor wireless scenario using the spectrum sharing scheme, the authors derive both upper bound and achievable bound of the transport capacity under practical deployment constraints. A test-bed is also developed to study the effect of the system parameters and validate the practicality of the proposed spectrum sharing concept. The experimental results demonstrate that the use of the IRS realizes a significant improvement on spectrum efficiency for the legacy transceivers. However, the test-bed only implements two transceivers. The practical performance of a large number of indoor communication pairs sharing the same spectrum band is still unknown.

\begin{figure}
\centering
\includegraphics[width=0.4\textwidth]{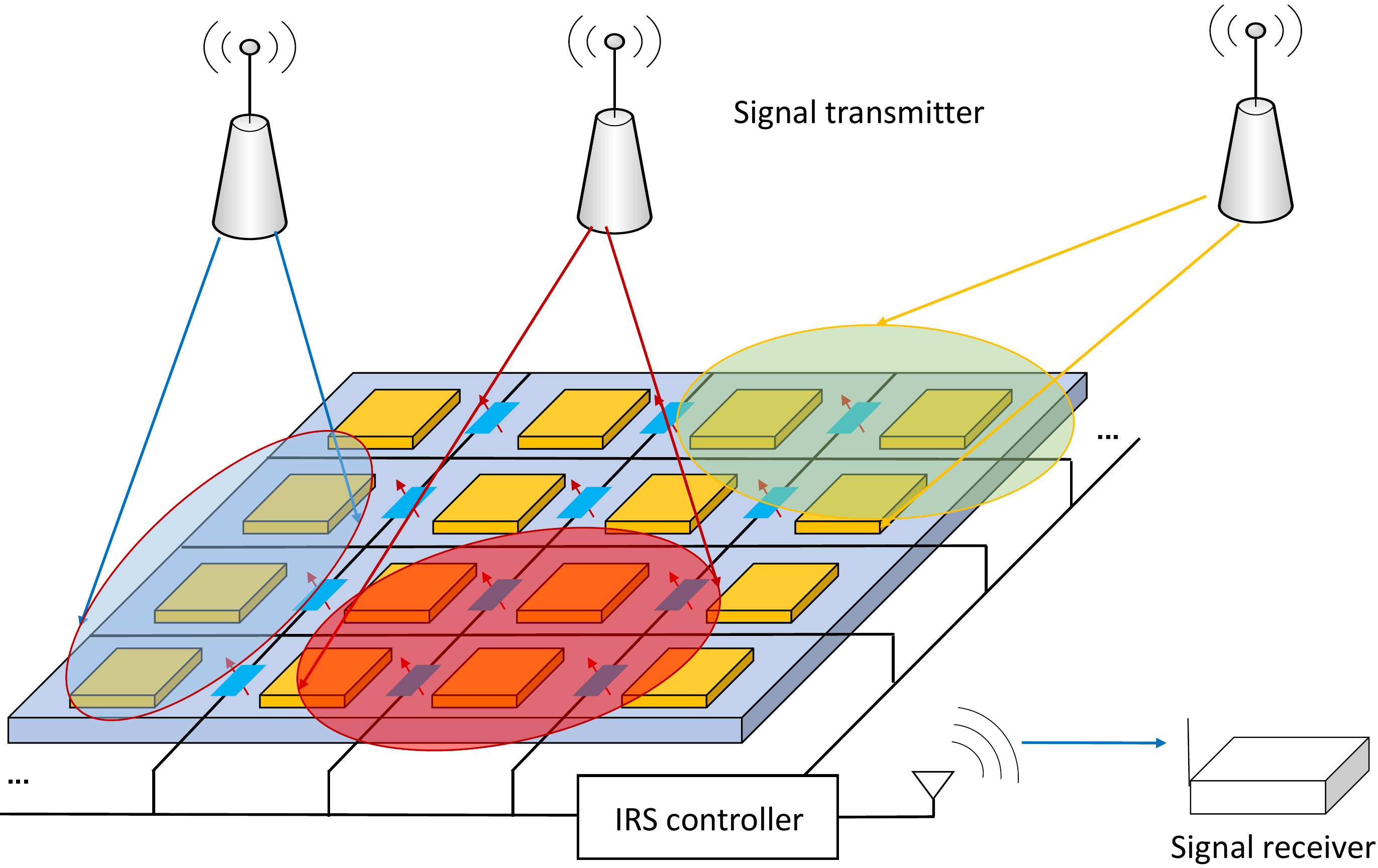}
\caption{The IRS is used as the signal receiver to enhance system capacity.} \label{fig_receiver}
\end{figure}

\subsection{Using IRS as the Signal Receiver}
Instead of using the IRS as a signal reflector for performance enhancement of wireless communications, the authors in~\cite{17capacity_sha} examine the performance of using the IRS as the signal receiver. In particular, the authors consider an uplink scenario where multiple single-antenna devices distributed in a three-dimensional (3D) space transmit to a vertical plane metasurface (e.g., hung on the wall), as shown in Fig.~\ref{fig_receiver}. Assuming large-size IRS and LOS channel conditions, the normalized capacity per unit of the IRS with perfect channel estimations can be characterized into an integral expression. The analytical expression demonstrates an asymptotic behavior that, given a constant transmit power, the normalized capacity approaches half of the transmit power divided by the spatial power spectral density of noise. The authors in~\cite{equalizer} consider a similar MU MIMO system, where the uplink transmissions to the IRS need to cancel interference among different users by optimizing the IRS's phase shifters. Based on properly design message passing schemes, the authors in~\cite{equalizer} propose decentralized algorithms for the reflecting element to iteratively update phase shifters simultaneously. However, both works in~\cite{17capacity_sha} and~\cite{equalizer} focus on idealistic models, assuming either perfect channel information or full control of the phase shifts.

Different from~\cite{17capacity_sha}, reference~\cite{19rate_walid} considers a more practical scenario where the IRS has a finite area and the interfering channels could either be LOS or non-LOS. With imperfect channel estimations, the authors derive the asymptotic capacity and reveal a channel hardening effect, i.e., the impacts of channel estimation errors and the non-LOS path are negligible when the number of scattering elements becomes large. The simulation results also demonstrate that, compared to conventional massive MIMO, a large-scale IRS can achieve a better reliability in terms of expectation and variance of capacity. Imperfect channel estimation is taken into account in~\cite{reliability}, where multiple IoT devices are connected to multiple non-overlapping IRS units in an IRS-assisted uplink system. The asymptotic sum-rate distribution with Rician fading can be derived in closed form and used for the analysis of outage probability. Numerical results verify that the IRS can provide reliable communications in terms of outage probability. The authors in~\cite{18capacity_degrade_sha} investigate the performance of IRS with hardware impairments which induce a higher effective noise level. A closed-form expression is derived to characterize the capacity of the uplink transmission to the IRS. To mitigate the negative effects of hardware impairments, the analytical and simulation results suggest a splitting of a large-scale IRS into an array of smaller IRS units. Different from the previous works focusing on capacity performance, reference~\cite{18positioning_sha} explores the potential of using an IRS with a large number of scattering elements for terminal positioning. The analysis also shows that deploying multiple IRSs in a distributed manner is an effective way to improve the coverage for terminal positioning.

\subsection{Using IRS as the Signal Transmitter}

Besides using IRS as a reflector or a receiver, it is also possible to use IRS as a source signal transmitter, similar to wireless backscatter communications that adapt the antenna's reflection coefficients by load modulation~\cite{irs_survey,backscatter18}. By controlling the phase shifts of the IRS's scattering elements, the outbound waves generated by the IRS can create different radiation patterns, which can carry information if these radiation patterns can be sensed and differentiated at the receiver. This is the main design principle of spatial modulation, which is conventionally realized by using reconfigurable antennas~\cite{indexmod17,17ee_renzo1,19spatial_renzo}. The IRS's superior reconfigurability and capability of reshaping the wireless environment make it much easier for its applications in spatial modulation~\cite{irs_survey}. Along this research direction, the authors in~\cite{irs_im19} design spatial modulation methods for the IRS, which can be realized by either creating exotic scattering pattern to enhance end-to-end transmissions or utilizing the index modulation of multiple receive antennas to render improved spectrum efficiency. Numerical results demonstrate that IRS-based spatial modulation methods provide a large capacity with ultra-low BER. The authors in~\cite{irs_survey} derive the error probability in closed form by using IRS to transmit M-ary phase modulated symbols. It shows that the IRS can transmit information with high reliability, similar to using it as a reflector.

The authors in~\cite{kim} use IRS to perform wireless backscatter communications leveraging its capability of manipulating the phases of the reflected signals. The authors in~\cite{channel-backscatter} similarly use an IRS as the backscatter transmitter and propose a general system model for the IRS-based ambient backscatter communications. Considering a large size of reflecting elements, we expect that the IRS-based backscatter communications will have higher flexibility than the conventional wireless backscatter communications, as the IRS is able to generate more exotic reflection patterns that can be used for information communications. This may lead to a higher data rate and larger transmission distance. In~\cite{pbit19}, the IRS is used to simultaneously enhance the legacy RF communications and send its private data to a receiver by spatial modulation. The information from both the RF transmitter and the IRS can be retrieved by a two-step decoding scheme. Similar to~\cite{pbit19}, the authors in~\cite{liangback} and~\cite{backlink} use IRS to perform backscatter communications to deliver its own data and simultaneously enhance the primary RF communications. An approximation is derived in~\cite{backlink} for the probability of the backscatter channel gain greater than the direct link, which is a useful performance measure to determine the number of the IRS's reflecting elements. The analysis and numerical results show that the use of IRS can achieve high-capacity wireless transmissions even without the LOS link. The authors in~\cite{sharing} integrate the IRS's backscatter communications into a cognitive radio network (CRN), promoting spectrum sharing between the secondary and primary users. The IRS on one hand assists the primary user's information transmissions, on the other hand backscatters its own information. The design objective is to maximize the secondary user's data rate, subject to a QoS requirement at the primary user, by jointly optimizing the secondary user's transmit power and the IRS's passive beamforming strategy in the alternating optimization framework. Simulation results show that the IRS-assisted CRN is efficient for secondary transmissions, even under a very challenging scenario where the secondary transmitter is much closer to the primary receiver.

\begin{table*}
\caption{Summary of Existing Literature in Performance Analysis of IRS Systems}
\begin{center}
\begin{tabular}{  l p{30mm} p{100mm} p{20mm}  p{15mm}  }
\hline
Reference &  Performance Metrics  &  System Model & Characterization  \\
\hline
\cite{19ref_prob_renzo}& Reflection prob. & Point-to-point communications in the presence of random metasurfaces  
& Exact    \\ \hline
\cite{18smart_mmwave} & Outage prob. &  Point-to-point mmWave communications with only reflection links  & Exact \\
\hline
\cite{17coverage} & Coverage prob.  & Downlink mmWave communications in the presence of LOS  blockages and reflectors & Approximation  \\ \hline
\cite{18errorpb_renzo1} & Bit error prob. & Point-to-point MIMO communications &  Exact    \\ \hline
\cite{ber-noma} & Bit error rate & MU MISO NOMA downlink  & Exact  \\ \hline
\cite{symbol-error} & Symbol error prob. & Point-to-Point SISO with Rayleigh fading channels & Upper bound   \\\hline
\cite{optimality} & Achievable rate & MU MISO downlink in the TDMA protocol with practical limitation of IRS & Asymptotic \\\hline
\cite{he2019adaptive} & Achievable rate & mmWave MIMO downlink systems with limited feedback from the mobile user & Asymptotic   \\\hline
\cite{rank-improve} &  Achieving rate &  Point-to-point MIMO system with rank improvement & Approximate   \\\hline
\cite{phaseshifts} & Achievable rate & IRS-assisted uplink system with finite resolution IRS & Approximate  \\\hline
\cite{spatial} & Spatial throughput & Single-cell MU downlink system with multiple randomly distributed IRSs &  Approximate\\\hline
\cite{capacity-region} & Capacity region & A multiple access channel for two users sending information to an AP & Upper bound    \\\hline
\cite{16sharing_capacity} & Transport capacity & Spectrum sharing among multiple transceivers through a reconfigurable metasurface & Upper bound   \\ \hline
\cite{17capacity_sha} & Normalized capacity   & Multiple devices in 3D space transmit to a vertical metasurface with LOS interference & Asymptotic   \\ \hline
~\cite{reliability} & Capacity, outage prob. &  MU uplink transmission to the IRS with imperfect channel estimation & Asymptotic  \\\hline
\cite{19rate_walid} & Ergodic capacity &  MU uplink transmission to a vertical metasurface with LOS and non-LOS interferences & Asymptotic      \\ \hline
\cite{18capacity_degrade_sha} & Ergodic capacity & A single-antenna user transmits to a vertical 2D metasurface with hardware impairments  & Upper bound \\
\hline
\cite{18positioning_sha} & Positioning coverage &   Positioning for a single-antenna device located in front of an IRS  & Lower bound   \\
\hline
\end{tabular}
\end{center} \label{Compare}
\end{table*}

{\bf Summary:} Table~\ref{Compare} summarizes and compares the existing literature reviewed in Section~\ref{sec_analysis}. The majority of these research works studies the performance of a standalone IRS system, i.e., either point-to-point communications or multiple communication pairs. References~\cite{19ref_prob_renzo} and~\cite{17coverage} analyze the performance of deploying the IRS reflectors in large-scale systems, however, under unrealistic assumptions such as the fixed-length model for random environmental reflectors~\cite{19ref_prob_renzo} and LOS links for all the reflectors~\cite{17coverage}. Besides, the channel information is mostly assumed to be known for the IRS's phase control, e.g.,~\cite{18errorpb_renzo1,ber-noma,symbol-error,optimality,he2019adaptive}. Hence, there is a need for more realistic models to analyze the performance of using the IRS in practical large-scale communications systems. Moreover, the above-mentioned literature hardly takes the user mobility into account, neither for indoor nor outdoor scenarios. User mobility introduces not only handoffs among different IRS units but also a spatial correlation in user distribution that may cause non-negligible impacts on the system performance. Hence, it becomes a critical research direction in the future to incorporate different mobility models into the performance analysis of IRS-assisted wireless systems. Additionally, the performance metrics currently under investigation are also limited in the literature. The theoretical performance of IRS-assisted wireless systems can be more thoroughly understood from the potential aspects as follows:
\begin{itemize}
\item {\em Pairwise error probability} which measures the probability that the decoded signal is a certain symbol given the transmitted signal.
\item {\em Average area spectral efficiency} which is the sum of the capacity of all the communication channels normalized over spectral bandwidth and spatial area.
\item {\em Energy efficiency} which measures the capacity normalized over the energy consumption of IRS systems.
\item {\em Handoff rate} which is the frequency of occurrences that a user handoffs to another IRS.
\end{itemize}

The stochastic performance analysis for IRS-assisted wireless systems mainly investigates the potential of the information-theoretic performance gain. For a specific network design problem, the maximization of performance gain requires a joint optimization of the active transceivers and the IRS's passive scattering elements. The joint phase control of scattering elements can be regarded as passive beamforming, which is closely coupled with the control variables of the active transceivers. This not only makes the performance analysis of IRS-assisted wireless systems more complicated, due to randomness and ubiquity of the scattering elements in the radio environment, but also results in new optimization problems that require novel solutions to account for the interactions between active and passive devices.

\section{Application and Optimization of IRS-Assisted Wireless Networks}\label{sec_optimization}

By smartly adjusting the phase shifts of all scattering elements, as illustrated in Fig.~\ref{fig:compare}, the reflected signals can be combined coherently at the intended receiver to improve the received strength or combined destructively at the non-intended receiver to mitigate interference. This can be verified by the experimental demonstration and channel measurements in~\cite{tang2019wireless}, which pave the way for further theoretical studies and system optimization. In the sequel, we review the main optimization formulations and solutions proposed for IRS-assisted wireless systems. The typical design objectives include SNR or rate maximization, transmit power minimization, EE/SE performance maximization, and physical layer security issues.

\subsection{SNR or Capacity Maximization}

\subsubsection{IRS-assisted Point-to-Point Communications}
Considering a point-to-point communication scenario, the authors in~\cite{18pbf_rui1} focus on an IRS-assisted multiple-input single-output (MISO) system, where one IRS with $N$ passive scattering elements is deployed to assist the downlink information transmission. A joint beamforming problem is formulated to maximize the received signal power at the user, by jointly optimizing the AP's transmit beamforming and the continuous phase shift of each scattering element. Semidefinite relaxation (SDR) is firstly proposed to obtain an approximate solution as a performance upper bound. Then, the alternating optimization is employed to update the active and passive beamforming strategies iteratively. Given the fixed passive beamforming, the AP's optimal beamforming is easily obtained by the maximum-ratio transmission strategy. Given the AP's beamforming, the IRS's optimal passive beamforming can be simply aligned with the direct channel to enhance the received signal power. Compared to the non-IRS-assisted MISO system, the SNR of the IRS-assisted MISO system can be improved by around 10 dB using an IRS with 100 scattering elements. Another important finding in~\cite{18pbf_rui1} is that the SNR at the receiver increases in the order of $N^2$. This power scaling law is further studied in~\cite{scaling_law} and compared with the massive MIMO system. Analytical results show that a large number of reflecting elements are required to obtain the SNR comparable to massive MIMO systems. The authors in~\cite{miso_robert} consider a similar IRS-assisted MISO downlink system. The joint optimization of the AP's transmit beamforming and the IRS's phase shifts is solved by fixed point iteration and manifold optimization techniques, respectively, which are shown to be effective in tackling the IRS's unit modulus constraints. These two algorithms not only achieve a higher data rate but also have a reduced computational complexity.

The above IRS-assisted MISO downlink model and the heuristic alternating optimization in~\cite{18pbf_rui1} provide a general framework for the optimized design of IRS-assisted systems, and thus can be extended to an upsurge of different network scenarios. The IRS-enhanced MISO OFDM downlink system is studied in~\cite{passive-ofdm}. To maximize the downlink achievable rate, a joint optimization of the BS's transmit power allocation and the IRS' s passive beamforming follows the alternating optimization framework. The MISO cognitive radio system is studied in~\cite{liangcrn}, where multiple IRSs are employed to maximize the data rate of the secondary receiver, subject to the interference constraints at the primary receivers. The authors in~\cite{mmwave-joint} also use multiple IRSs to assist mmWave MISO communications. Through joint active and passive beamforming optimization, the IRS can provide enhanced paths of reflection for mmWave signals, and thus can maximize the received signal power and extend the network coverage. Both of the phase shifts and transmit beamforming are derived optimally in closed forms for the single-IRS case. This is achieved by exploiting the characteristics of mmWave channels, i.e., assuming a rank-one channel matrix between the BS and the IRS. The analysis and simulation results reveal that the received signal power increases quadratically with the number of reflecting elements, which verifies the power scaling law revealed in~\cite{18pbf_rui1}.

Different from the MISO systems in~\cite{miso_robert,passive-ofdm,mmwave-joint}, the authors in~\cite{capacity_rui} characterize the capacity limit of an IRS-assisted MIMO system, by jointly optimizing the IRS's reflection coefficients and the MIMO transmit covariance matrix. The capacity maximization for broadband transmissions is considered in frequency-selective fading channels, where transmit covariance matrices can be optimized for different OFDM sub-carriers. Based on convex relaxation, the alternating optimization algorithm used in~\cite{18pbf_rui1} is modified to find a high-quality sub-optimal solution. Numerical results show that the IRS-assisted MIMO system achieves substantial capacity improvement compared to traditional MIMO systems without IRS, e.g., the improvement is over 45\% in the high SNR regime when using an IRS with 80 scattering elements. The point-to-point mmWave MIMO OFDM system is studied in~\cite{gmd-hybrid}. The hybrid MIMO beamforming matrices and the IRS's phase shift matrix are separately optimized to achieve better BER and spectrum efficiency performance than the conventional approaches. Considering a similar IRS-assisted MIMO system, the authors in~\cite{ergodic} derive an upper bound for the ergodic capacity with Rician fading channel. To maximize the capacity upper bound, the IRS's phase shift matrix is optimized by SDR and Gaussian randomization methods. In~\cite{sumpath}, the data rate maximization of an IRS-assisted MIMO system is formulated into a mixed integer problem. Based on the alternating optimization framework, the ADMM algorithm is leveraged to find the phase shifts of individual scattering elements, and then the active beamforming is obtained by classic singular value decomposition and water-filling solutions. Focusing on a two-way full-duplex IRS-assisted MIMO system, the authors in~\cite{two-way} maximize the sum rate by a joint optimization of the precoders at two transmitters and the IRS's phase shift matrix. The iterative solution follows a similar alternating optimization framework as that in~\cite{18pbf_rui1,capacity_rui}. Comparing to conventional relay-assisted full-duplex MIMO system, the performance of the IRS-assisted system is comparable to that of a relay with transmit power at around $-40$ dBm to $-35$ dBm. We note that all aforementioned research works in~\cite{miso_robert,passive-ofdm,mmwave-joint,capacity_rui,ergodic,gmd-hybrid,two-way} rely on the ideally designed IRS with infinite phase resolution, i.e., the phase shift of each scattering element can be fully controllable. However, this is difficult for practical implementation and also complicated for designing exact phase control algorithms. Moreover, full CSI is generally required for the IRS controller to make perfect phase control. This implies that the overhead of information exchange can be prohibitively high, especially for self-sustainable IRS via wireless energy harvesting.

Ideally, the reflection amplitude is constant and assumed to be independent of the phase shift. Based on experimental results, the authors in~\cite{shift-model} notice the non-ideality in controlling the IRS's phase shifts, and thus propose a practical phase shift model that captures the nonlinear dependence between phase and amplitude. Employing this new model in an IRS-assisted MISO system, a joint beamforming design problem is formulated to maximize the achievable rate in MISO downlink transmissions, which can be solved in the alternating optimization framework. The capacity degradation is evaluated in~\cite{stat-csi} by considering different levels of quantization in the IRS's discrete phase control. Based on statistical CSI, a tight approximation of the ergodic capacity is derived for an IRS-assisted MISO system, and maximized by optimizing the IRS's phase shift matrix. Numerical results show that a 2-bit phase quantizer is sufficient to ensure capacity degradation of no more than 1 bit/s/Hz. This greatly simplifies the practical implementation, design costs, and applications of IRS in wireless communications. In~\cite{liangcrn}, the imperfect CSI is modeled by a norm-based uncertainty set and a worst-case robust optimization is formulated to optimize the joint active and passive beamforming strategy. Considering the channel estimation errors and training overhead, the authors in~\cite{you2019intelligent} formulate an optimization problem to maximize the achievable data rate by designing the IRS's discrete phase tuning strategy. A low-complexity successive refinement algorithm is devised to achieve a high-quality sub-optimal solution with proper algorithm initialization. Similarly, considering discrete phase shift and unknown CSI in~\cite{discrete-shift}, a tight lower bound is derived for the user's asymptotic rate in IRS-assisted MISO downlink communications. Numerical results reveal the discrete phase design with moderate to high phase resolutions can asymptotically approach that of the optimal continuous phase control with perfect CSI.

To reduce the overhead in channel estimation, the authors in~\cite{irs_ofdm} propose an element grouping method to exploit the channel spatial correlation in an IRS-assisted OFDM system. By estimating the combined channel of each group, the training overhead can be substantially reduced. The alternating optimization method is also used to maximize the achievable rate by optimizing the BS's power allocation and the IRS's passive beamforming with a customized algorithm initialization. A significant performance improvement on the link rate can be observed compared to the cases without IRS. The authors also show that there exists an optimal size for grouping to maximize the achievable rate. The authors in~\cite{phase-error} focus on the phase error in channel estimation, which brings the ambiguity in the IRS's phase tuning. The composite channel assisted by an IRS with $N$ reflecting elements can be modeled as Nakagami fading channel. Theoretical analysis reveals that the average received SNR still grows with $N^2$ and the error probability performance is robust against the phase errors. The authors in~\cite{ss_asym} quantify the negative effect of pilot contamination in channel estimation due to intra/inter-IRS interference. An asymptotical analysis on the spectrum efficiency of multi-IRS-assisted uplink transmissions reveals that the achievable spectrum efficiency is limited by the effect of pilot contamination and intra/inter-IRS interference even with infinite reflecting elements. The above analyses with finite phase resolution~\cite{stat-csi}, inexact CSI~\cite{discrete-shift,you2019intelligent,irs_ofdm}, estimation and quantization errors~\cite{phase-error,ss_asym} verify that the practical imperfections can be generally counteracted by using a larger-size IRS. The previous analyses in~\cite{reliability,18capacity_degrade_sha} reveal another countermeasure, i.e., dividing larger IRS into multiple smaller IRSs, to enhance reliability and fight against channel estimation errors. All these studies provide useful guidelines for the IRS's practical implementations in wireless systems.

\subsubsection{Multi-user or Multi-cell Coordinated Communications}
The previous part reviews and verifies the potential performance enhancement of point-to-point wireless communications by deploying IRS in the propagation environment. The SNR or capacity maximization problems can be naturally extended to the MU scenarios. However, the solution methods will become more involved due to the resource competition and interference among different users.

The authors in~\cite{18maxrate_chau} present an efficient design for sum-rate maximization in IRS-assisted downlink MISO communications, subject to the AP's power budget constraint. Each user is given a minimum data rate requirement. The sum-rate maximization problem is firstly simplified by using the zero-forcing (ZF) transmission scheme and then following an iterative procedure to optimize the transmit power and the phase shift matrix. This is achieved by combining the alternating maximization with the majorization-minimization (MM) method. The sum-rate maximization in a similar model is also studied in~\cite{zhao-icc} and solved heuristically by the alternating optimization framework. In the ideal case with infinite phase resolution, the authors in~\cite{18maxrate_chau} show that the system throughput can be increased by at least 40\%, without additional energy consumption. The optimal solution to the sum-rate maximization relies on the exact CSI, which is difficult to obtain for large-size IRS. Considering a similar IRS-assisted MU MISO system, the authors in~\cite{weighted-twc} extend the sum-rate maximization problem with perfect CSI to the case with imperfect CSI. For perfect CSI case, the active and passive beamforming strategies can be obtained by using the fractional programming technique, which can be extended to the case with imperfect CSI based on a stochastic successive convex approximation (SCA) method. Simulation results reveal that the proposed algorithm performs quite well when the channel uncertainty is smaller than 10\%. To avoid frequent channel estimation, the authors in~\cite{two-timescale} propose a two-timescale algorithm to maximize the sum-rate for an IRS-assisted MU MISO downlink system. The IRS's phase shift matrix is firstly optimized based on the slow-varying statistical CSI, while the AP's transmit beamforming is then optimized according to the instantaneous CSI. Besides, different from the popular SDR-based algorithms~\cite{18pbf_rui1}, the authors in~\cite{two-timescale} leverage the penalty dual decomposition technique that allows parallel updates of the optimization variables. Hence, the proposed algorithm can significantly reduce the channel training overhead, computational time and complexity for optimizing the IRS's passive beamforming. Our previous observation in~\cite{spatial,capacity-region} reveals that the capacity gain using IRS becomes more significant for asymmetric user rates. In particular, the capacity gain can be achieved by deploying IRS to assist NOMA transmissions to cell-edge users simultaneously with the other users closer to the BS~\cite{irs-noma}. Hence, we expect that the IRS-assisted sum-rate maximization will greatly improve the users with weak channel conditions. However, it usually leads to a non-convex design objective due to the interference and couplings among different users.

The design objectives of sum-rate maximization in different network scenarios are intrinsically fighting against the resource competition or interference among different users. The use of the IRS can make the wireless propagation channels more flexible to control and thus easier for interference suppression. Considering a typical MU MIMO system, the authors in~\cite{program-channel} show the feasibility of constructing multiple interference-free beams by using the IRS with a large number of passive elements. Analysis shows that a single set of optimal beamforming weights can form multiple interference-free beams for multi-stream MIMO transmissions. The authors in~\cite{multigroup} investigate users' interference in a more complicated multi-cell multi-cast downlink system. The information destined to different groups of users are independent and different. Hence, there exists inter-group and inter-cell interference. To maximize the sum-rate of all groups, a concave lower bound of the objective function is firstly derived and then the alternating optimization is employed to update iteratively the BS's precoding matrix and the IRS's passive beamforming. To reduce the computational complexity, the MM method in~\cite{18maxrate_chau} is adopted to derive a closed-form solution in every iteration. The simulation results demonstrate that the sum-rate can be improved by more than 100\% when assisted by an IRS with only 8 scattering elements, comparing to a massive MIMO system with 256 antennas at the BS. To mitigate inter-cell interference, the authors in~\cite{multicell-relaying} deploy the IRS at the cell boundary of a multi-cell system to assist the downlink MIMO transmissions to cell-edge users. The maximization of weighted sum-rate is solved with a similar alternating optimization as that in~\cite{multigroup}. The BS's active precoding and the IRS's passive beamforming are iteratively optimized by using the block coordinate descent (BCD) algorithm. Note that it is difficult to guarantee individual users' rate requirements in a sum-rate maximization problem. Though we can add individuals' rate constraints as that in~\cite{18maxrate_chau,zhao-icc}, an improper or uneducated setting for the users' minimum rate requirement may drastically bring down the sum-rate performance, or even make the design problem infeasible.

The difficulties in sum-rate maximization can be resolved by considering user fairness as the critical performance metric for performance maximization in MU networks, which in general can be formulated as max-min problems. The authors in~\cite{multiaccess-noma} maximize the minimum signal-to-interference plus noise ratio (SINR) of all users in an IRS-assisted NOMA system by jointly optimizing the BS's power allocation and the IRS's passive beamforming. Simulation results show that the IRS with 1-bit phase resolution improves the max-min rate by 20\% compared to that of traditional NOMA. Similarly, the authors in~\cite{17maxrate_sha} maximize the minimum user-rate in the wireless system assisted by a set of distributed IRS units. Each IRS unit has a separate signal processing unit and is connected to a central processing unit that coordinates the behaviors of all the IRS units. A user assignment scheme between each user and the IRS is proposed to improve the minimum user-rate. The optimal user assignment scheme can be effectively found by solving classical linear assignment problems defined on a bipartite graph. Numerical results show that the proposed user assignment scheme is close to optimum both under LOS and scattering environments. The maximization of the minimum weighted SINR in a multi-cell MISO downlink system is studied in~\cite{fariness-multicell}, where an IRS is dedicatedly deployed to suppress inter-cell interference and assist information transmission of cell-edge users. With the fixed passive beamforming at the IRS, the BSs' transmit powers are optimized by second-order-cone programming (SOCP). The update of the IRS's passive beamforming can be achieved by using SDR and SCA methods. Numerical results show that the proposed algorithm can achieve significant performance gain, i.e., the minimum SINR can be increased by over 150\% with the BS's transmit power at $35$ dBm, compared to the conventional case without IRS. The common observation from the above works in~\cite{18maxrate_chau,zhao-icc,two-timescale,program-channel,multigroup,irs-noma,multicell-relaying,multiaccess-noma,17maxrate_sha,fariness-multicell} is that they all consider interference constrained systems, where the spectrum access of different users introduces mutual interference to each other and thus limits the sum-rate performance. This also complicates the algorithm design due to the users' couplings.

\begin{table*}[t]
\caption{SNR or Capacity Maximization in IRS-assisted Wireless Networks}
\begin{center}
\begin{tabular}{  l p{18mm} p{20mm} p{10mm}  p{30mm} p{60mm}  }
\hline
Reference &  Optimization & System Model & Resolution & Design Variables  & Main Results \\
\hline
\cite{18pbf_rui1}& Max. signal power & MISO downlink  & Continuous & Transmit and passive beamforming & The IRS with size $N$ achieves a total beamforming gain of $N^2$ \\ \hline
\cite{miso_robert} & Max. data rate &  MISO downlink & Continuous &  Transmit and passive beamforming & Achieve a higher data rate with reduced computational complexity \\ \hline
\cite{passive-ofdm} & Max. data rate & MISO OFDM downlink & Continuous &  Transmit power allocation, passive beamforming & Effective in boosting the data rate of a cell-edge user \\\hline
\cite{liangcrn} & Max. data rate & MISO CRN downlink & Continuous &  Transmit and passive beamforming & IRSs can improve the data rate of the secondary user under both perfect and imperfect CSI cases \\\hline
\cite{mmwave-joint} & Max. signal power  &  MISO mmWave downlink  & Continuous & Transmit and passive beamforming & The received signal power increases quadratically with $N$  \\\hline
\cite{capacity_rui} & Max. capacity &  MIMO downlink  & Continuous & Transmit covariance matrix, passive beamforming & Substantially increased capacity compared to MIMO channels without using the IRS \\\hline
\cite{gmd-hybrid} & Max. data rate & mmWave hybrid MIMO OFDM & Continuous & Hybrid precoders, passive beamforming & Achieve better BER and spectrum efficiency than the conventional approaches \\\hline
\cite{ergodic} & Max. capacity & MIMO downlink & Continuous & Passive beamforming & Capacity gain validated by simulations \\\hline
\cite{two-way} & Max. sum rate & Two-way full-duplex MIMO & Discrete & Transmitters' precoders, passive beamforming & Capacity gain is close to that of a full-duplex relay with transmit power at $-40\thicksim-35$ dBm \\\hline
\cite{stat-csi} & Max. capacity & MISO downlink, statistical CSI & Discrete & Discrete phase shifts &  2-bit phase quantizer is sufficient to ensure capacity degradation of no more than 1 bit/s/Hz \\\hline
\cite{you2019intelligent} & Max. data rate & SISO uplink & Discrete & Discrete phase control & Significant rate improvement is achieved by a low-complexity successive refinement algorithm \\\hline
\cite{discrete-shift} & Max. data rate & MISO downlink, unknown CSI& Discrete & Discrete phase shifts & Moderate to high phase resolution approximates the optimum with continuous phase and perfect CSI\\\hline
\cite{irs_ofdm} & Max. data rate &  SISO OFDM downlink   & Continuous & Transmit power allocation, passive beamforming & An optimal grouping size exists to maximize the achievable rate\\ \hline
\cite{18maxrate_chau} & Max. sum-rate & MU MISO downlink & Continuous & Transmit power allocation, passive beamforming & Throughput increased by at least 40\%, without requiring additional energy consumption \\\hline
\cite{zhao-icc} & Max. sum rate & MU MISO downlink  & Continuous & Transmit power allocation, passive beamforming & Significant performance gain achieved over the random phase shift and the conventional ZF methods \\\hline
\cite{two-timescale} & Max. sum rate & MU MISO with statistical CSI & Discrete & Transmit precoders, discrete phase shifts & Significantly reduce the channel training overhead, computational time and complexity for optimizing the IRS's passive beamforming \\\hline
\cite{irs-noma} & Max. sum rate & MU MISO NOMA downlink & Discrete, binary & Discrete phase shifts & Performance gain of IRS-assisted NOMA becomes significant for asymmetric user rates \\\hline
\cite{program-channel} & Min. interference & MU MIMO downlink & Continuous & Passive beamforming & A single set of weights can form multiple interference-free beams for MIMO transmission  \\\hline
\cite{multigroup} & Max. sum-rate & MU multi-cell MISO downlink & Continuous & Precoding matrix, passive beamforming & Improved EE/SE performance over conventional massive MIMO systems  \\\hline
\cite{multicell-relaying} & Max. weighted sum-rate & MU multi-cell MIMO downlink & Continuous & Precoding matrix, passive beamforming & Significant performance gain is achieved over the conventional counterpart without the IRS  \\\hline
\cite{multiaccess-noma} & Max. min SNR & MU SISO NOMA downlink & Continuous & Power allocation, passive beamforming   & 1-bit phase resolution improves the min rate by 20\% compared to that of traditional NOMA  \\\hline
\cite{17maxrate_sha} & Max. min rate &  MU uplink  & Continuous & User-IRS assignment, passive beamforming   &  The user assignment performs close to the optimum in both LOS and scattering environments \\\hline
\cite{fariness-multicell} & Max. min SINR & Multi-cell MISO downlink & Continuous & Transmit and passive beamforming & Min SINR can be increased by over 150\% with the BS's transmit power at $35$ dBm \\\hline
\cite{symbol_precoding} & Min. worst-case SER & MU MIMO downlink & Discrete, binary & Precoding matrix, discrete phase control & Enhanced SER performance can be achieved \\\hline
\cite{guo2019weighted} & Max. weighted sum-rate & MU MISO downlink & Discrete & Transmit beamforming, discrete phase control & The IRS with 2-bit resolution achieves sufficient capacity gain \\\hline
\cite{di2019hybrid} & Max. sum-rate & MU MISO downlink & Discrete &  Transmit and passive beamforming  & A good performance gain achieved by using the IRS with a few discrete phase shifts\\ \hline
\cite{ofdma} & Max. min rate & MU OFDMA downlink & Continuous & RB and power allocation, passive beamforming &  The IRS-assisted system achieves massive MIMO gains with a fewer number of AP's antennas \\\hline
\end{tabular}
\end{center} \label{tab:capacity}
\end{table*}

Different from the above interference-limited networks, the authors in~\cite{ofdma} integrate IRS to an orthogonal frequency division multiple access (OFDMA) based MU downlink system. A joint optimization of the IRS's passive beamforming and OFDMA resource block (RB) as well as power allocations is proposed to maximize the minimum user rate. By using a dynamic passive beamforming scheme, the IRS's reflection coefficients are allowed to dynamically change over different time slots. In each time slot, only a subset of the users will be selected and served simultaneously, thus achieving a higher passive beamforming gain. Numerical results show that the dynamic passive beamforming outperforms the fixed passive beamforming scheme. The performance improvement becomes larger as the size of scattering elements increases. To this point, all the above works assume continuously controllable phase variables at the IRS. The analysis of performance degradation with practical low-resolution IRS is desirable for evaluating the robustness and reliability in practice.

The authors in~\cite{symbol_precoding} use a low-resolution IRS and design a symbol-level precoding scheme for MU MISO/MIMO downlink system to minimize the worst-case symbol-error-rate (SER). The discrete phase shifts are firstly relaxed as continuous design variables and optimized by the Riemannian conjugate gradient algorithm. Then, the low-resolution precoding vector is obtained by direct quantization. As a special case, the branch-and-bound method is proposed to reduce the quantization error for the 1-bit symbol-level precoding vector. Similarly, the authors in~\cite{di2019hybrid} employ the low-resolution IRS to enhance the sum-rate in MU MISO downlink system. A good sum-rate performance gain can be achieved by using the IRS with a reasonable size and a small number of discrete phase shifts. Considering weighted sum-rate maximization in a similar MU MISO system, the authors in~\cite{guo2019weighted} propose three iterative methods to optimize the discrete phase levels of the IRS's scattering elements in addition to the BS's beamforming optimization, based on the fractional programming method. Numerical results show that the IRS with 2-bit phase resolution achieves sufficient capacity gain with only a small performance degradation. Similar observations and implications are also made in~\cite{stat-csi,discrete-shift}. Note that the aforementioned works all assume perfect or static CSI of all users. In fact, the channel estimation involving passive IRS becomes more complicated than before due to a large size of passive reflecting elements. Besides, the channel estimation in a dynamic environment is inevitably contaminated by error estimates and thus leading to channel uncertainty, which has not been fully investigated for the IRS-assisted MU scenarios in the current literature.

\begin{figure*}[t]
	\centering
	\includegraphics[width=0.7\linewidth]{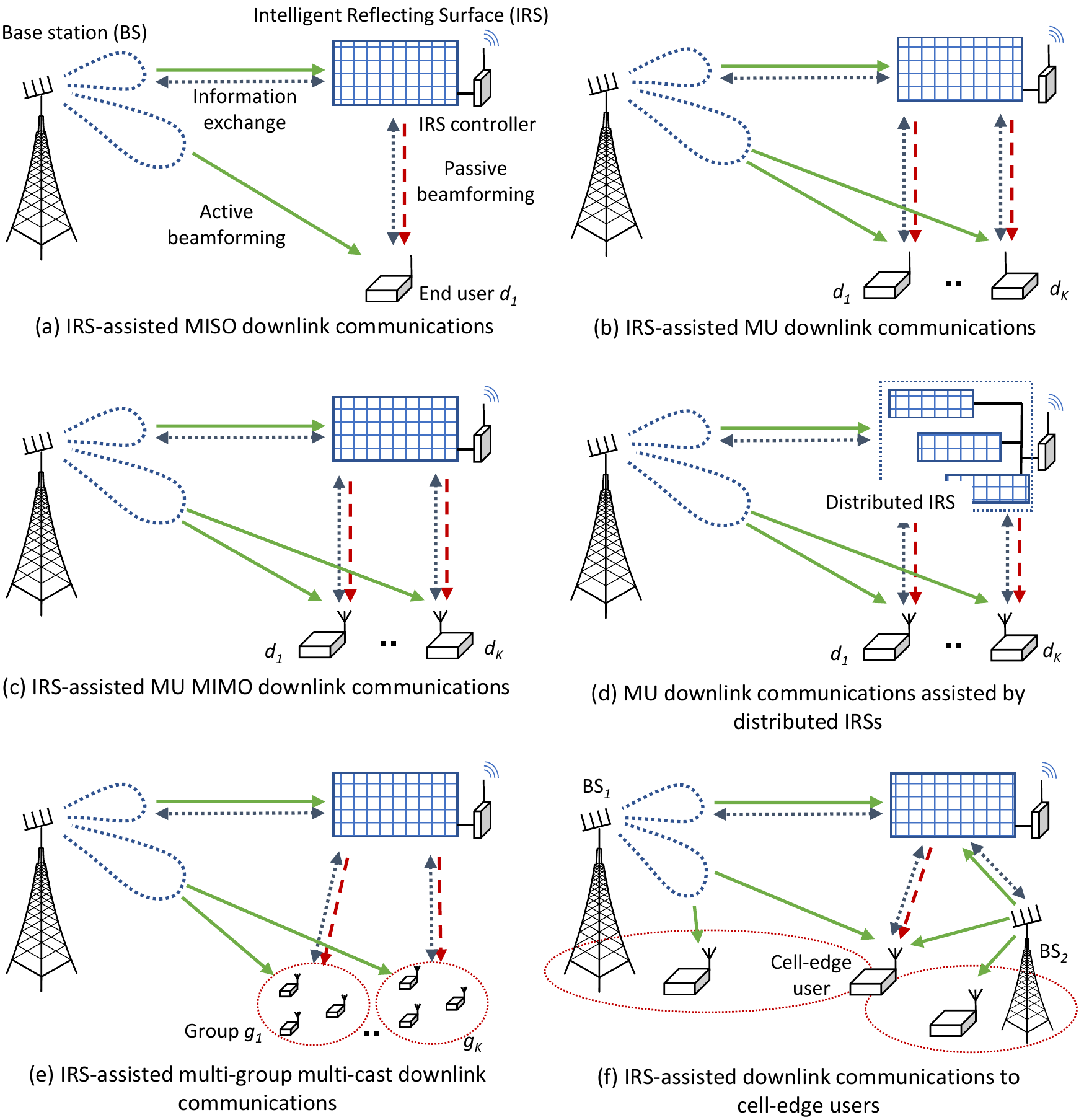}
	\caption{Different system models for IRS-assisted wireless networks. (a) Basic model of IRS-assisted downlink transmission from multi-antenna BS to one receiver, which can be a single-antenna end user, e.g.,~\cite{19ee_chau,18pbf_rui1,passive-ofdm,mmwave-joint,19pbf_rui2,18pbf_rui2,19minfu_noma,miso_robert,stat-csi,discrete-shift}, or equipped with multiple receiving antennas, e.g.,~\cite{capacity_rui,irs_ofdm,gmd-hybrid,ergodic}. (b) The IRS-assisted MU MISO downlink transmissions, e.g.,~\cite{18maxrate_chau,zhao-icc,two-timescale,irs-noma,ofdma,guo2019weighted,han2019intelligent,di2019hybrid,19ee_chau2}. The special case with single-antenna BS is studied in~\cite{irs_ofdm,multiaccess-noma,noma-oma}. (c) The IRS-assisted downlink transmissions to multi-antenna receivers (MIMO), e.g.,~\cite{symbol_precoding,program-channel}. (d) Multiple-IRS-assisted MISO/MIMO downlink transmissions, e.g.,~\cite{spatial,reliability,18capacity_degrade_sha,mmwave-joint,17maxrate_sha,liangcrn}. (e) IRS-assisted multi-group multi-cast downlink transmissions, e.g.,~\cite{multigroup,bf_multicluster}. (f) IRS-assisted downlink transmissions to cell-edge users, e.g.,~\cite{fariness-multicell,multicell-relaying}. }
	\label{fig:irs-assisted-model}
\end{figure*}

{\bf Summary}: In this part, we have reviewed the potentials of using the IRS to improve transmission performance in terms of SNR or data rate for both point-to-point communications and multi-group/multi-cell MU cases. A summary of existing works on SNR or capacity maximization is listed in Table~\ref{tab:capacity}. In particular, we have discussed the applications of the IRS under different communication models, e.g.,~MISO, MIMO, OFDM, NOMA, mmWave, and multi-cell systems. Different system models are illustrated and compared in Fig.~\ref{fig:irs-assisted-model}. The performance maximization of IRS-assisted wireless systems in different scenarios is typically formulated into a joint optimization problem of the IRS's passive beamforming and the BS's transmit beamforming or power allocation strategy. Along this main line of research, some special cases are also discussed, including the phase shift optimization for the non-ideal IRS with low phase resolution, or with incomplete or uncertain channel information. We notice that the main solution methods are based on a simple alternating optimization framework, which can guarantee the convergence to sub-optimal solutions. However, comparing the optimum, the performance loss by using the alternating optimization is not known exactly and seldom characterized in literature. By developing more sophisticated algorithms in the future work, we envision that the IRS-assisted wireless systems can achieve a higher performance gain than that in the current literature.

\subsection{Power Minimization or EE/SE Maximization}

Besides SNR and rate maximization, the IRS-assisted wireless networks can also help minimize the BS's transmit power or maximize the overall EE/SE performance. The IRS can configure wireless channels in favor of information transmission between transceivers. This results in a more energy-efficient communication paradigm, e.g., the BS can maintain the same transmission performance with a reduced power consumption. As such, IRS-assisted communications can be envisioned as a green technology for future wireless networks.

Focusing on an IRS-assisted MISO downlink scenario as that in~\cite{18pbf_rui1}, the authors in~\cite{18pbf_rui2} minimize the AP's transmit power under individual users' SINR constraints by jointly optimizing the AP's transmit beamforming and the IRS's passive beamforming strategies. Following SDR procedure and alternating optimization, the AP's transmit beamforming can be efficiently optimized by solving SOCP, and the optimization of the IRS's passive beamforming is degenerated to a conventional relay beamforming optimization problem. By an asymptotic analysis with an infinite number of scattering elements, the AP's transmit power can be scaled down in the order of $1/N^2$ without compromising the SNR at the receiver. Numerical results verify that the AP's transmit power can be reduced by more than 55\% for the users far away from the AP (e.g.,~50 meters). The authors in~\cite{han2019intelligent} focus on an IRS-assisted MISO broadcasting system and derive a lower bound of the BS's minimum transmit power, which is much lower than the cases without IRS. Moreover, the BS's transmit power can approach the lower bound as the number of scattering elements increases. Transmit power minimization is also considered in an IRS-assisted NOMA downlink system~\cite{19minfu_noma}. The joint optimization the active and passive beamforming is a highly intractable bi-quadratic program, which is firstly relaxed via SDR and then solved by the difference-of-convex (DC) algorithm. Simulation results demonstrate that the AP's transmit power can be reduced by more than 8 dB when using an IRS with 50 scattering elements. It is obvious that~\cite{18pbf_rui2,han2019intelligent,19minfu_noma} focus on point-to-point scenarios with simplified assumptions, e.g.,~fully controllable phase shifts, static and known CSI. The proposed design problems are expected to consider more general cases with multiple users in a dynamic environment.

The transmit power minimization for MU MISO downlink system is investigated in~\cite{bf_multicluster}, where the users are grouped into different clusters and NOMA is employed in each cluster to enhance information transmission. An effective SOCP-based alternating direction method of multipliers (ADMM) is proposed to optimize the BS's transmit beamforming and the IRS's passive beamforming. A ZF-based sub-optimal algorithm is also proposed to reduce the computational complexity. The simulation results demonstrate significant performance gain over the conventional SDR-based algorithms, e.g.,~\cite{18pbf_rui1,ergodic,fariness-multicell,19minfu_noma}. To compare multiple access scheme for MU downlink systems, the authors in~\cite{noma-oma} evaluate the minimum transmit powers required by different schemes. Assuming perfect CSI, the transmit power in either scheme is minimized by alternating optimization. The comparison reveals that the IRS-assisted NOMA may perform worse than that of the IRS-assisted time division multiple access (TDMA) for users closer to the IRS. To achieve NOMA gain over TDMA, it is preferred to pair users with asymmetric rates for NOMA downlink transmissions. Note that the BS in MISO downlink systems typically has constant power supply. Hence, the minimization of BS's transmit power may not be an urgent need. In fact, the BS's energy efficiency can be a more critical design objective, which represents the transmission capability of the network.

Instead of minimizing transmit power, the authors in~\cite{19ee_chau} maximize the energy efficiency by jointly optimizing the IRS's passive beamforming and the AP's power allocation over different users, subject to the AP's maximum power and the users' minimum QoS constraints. The total power consumption includes the transmit power, the constant circuit power, as well as the IRS's power consumption, which relates to the size and implementation of the reflecting elements. In particular, a finer phase resolution or a larger size of scattering elements implies a higher power consumption. Let $P_{IRS}$ denote the IRS's power consumption. The authors in~\cite{19ee_chau} present a simple linear power consumption model, i.e.,~$P_{IRS} = NP(b)$, where $N$ denotes the number of scattering elements and $P(b)$ is the power consumption of each phase shifter with $b$-bit resolution. Typical power consumption values of $P(b)$ are 1.5, 4.5, 6, and 7.8 mW for 3-, 4-, 5-, and 6-bit resolution phase shifting, respectively. A gradient descent method is firstly used for optimizing the IRS's phase control, and then the transmit power allocation is optimized by a fractional programming method. The simulation results in a realistic environment show that the IRS-assisted system can provide up to 300\% higher energy efficiency than that of the conventional multi-antenna AF-relay communications. The authors in~\cite{irs-df} compare the performance between the IRS-assisted system and the conventional decode-and-forward (DF) relay communications. The achievable rates for both cases are maximized by optimizing the transmit power and the size of the IRS. The main observation is that a large-size IRS is needed to achieve better performance than that of DF-relay communications, in terms of EE/SE performance. Though a power consumption model is proposed for the IRS, the IRS's power budget constraint is not considered in the optimization framework. Instead, the IRS is implicitly assumed to be always online and capable of phase tuning.

\begin{table*}
\caption{Power Minimization or EE Maximization in IRS-assisted Wireless Networks}
\begin{center}
\begin{tabular}{  l p{20mm} p{20mm} p{10mm}  p{30mm} p{60mm}  }\hline
Reference &  Optimization & System Model & Resolution & Design Variables  & Main Results \\\hline
\cite{18pbf_rui2} & Min. transmit power & MISO downlink & Continuous  & Transmit and passive beamforming &  Transmit power can be scaled down in the order of $1/N^2$ without compromising the SNR at the receiver \\\hline
\cite{han2019intelligent} & Min. transmit power & MISO downlink broadcasting & Continuous  & Transmit and passive beamforming & A lower bound is derived for the BS's minimum transmit power  \\\hline
\cite{19minfu_noma} & Min. transmit power & MISO NOMA downlink & Continuous  & Transmit and passive beamforming &  Verified effectiveness and superiority of using IRS to reduce the total transmit power \\\hline
\cite{bf_multicluster} & Min. transmit power &  MISO NOMA downlink & Continuous & Transmit and passive beamforming & The IRS-assisted ZF scheme outperforms the SDR-based algorithm when $N$ is large  \\\hline
\cite{noma-oma} & Min. transmit power & MU SISO NOMA downlink & Discrete & Transmit power allocation, discrete phase shifts & NOMA gain can be achieved by pairing users with asymmetric rates \\\hline
\cite{19ee_chau} & Max. energy efficiency & MISO downlink & Continuous  & Power allocation, passive beamforming & IRS-assisted system provides up to 300\% increase in energy efficiency than AF relay communications \\\hline
\cite{19pbf_rui2} & Min. transmit power &  MISO downlink & Discrete  & Transmit beamforming, discrete phase control & A discrete IRS can achieve the same power gain as that with infinite phase resolution \\ \hline
\cite{19ee_chau2} &  Max. energy efficiency &  MU MISO downlink & Discrete, binary & Transmit beamforming, discrete phase control & 1-bit resolution IRS significantly improves energy efficiency compared to relay-assisted communications \\ \hline
\end{tabular}
\end{center} \label{system-model}
\end{table*}

Prior works mostly assume infinite phase resolution for the IRS, which however is practically difficult to realize due to the hardware limitation. Considering a more practical case, the authors in~\cite{19pbf_rui2} and~\cite{discrete-tcom} minimize the AP's transmit power in a downlink MISO system, assisted by an IRS with finite phase resolution. The feasible discrete phase shifts can be obtained by quantization projection from the optimized continuous phase values, similar to~\cite{symbol_precoding}. Analytical results show that a practical IRS with discrete phase shifts can still achieve the same power scaling law as that with continuous phase shifts, the finding similar to~\cite{18pbf_rui1}. More interestingly, the performance loss due to quantization errors is shown to be irrelevant to $N$, the number of reflecting elements, while only dependent on the IRS's phase quantization level $b$. Numerical results reveal that the discrete phase shifts with $b=2$ or $b=3$ are sufficient to achieve the close-to-optimal performance, which is consistent with the former observation in~\cite{stat-csi,discrete-shift,guo2019weighted}. The optimization solutions to the single-user MISO system in~\cite{19pbf_rui2} can be extended to an MU case in~\cite{discrete-tcom}. However, the quantization projection used in~\cite{19pbf_rui2} may lead to higher performance degradation in the MU case due to the severe co-channel interference. A special case with the 1-bit phase resolution is studied in~\cite{19ee_chau2}. The optimization of phase shift matrix and power allocation for each user follows a similar alternating optimization method. The IRS's perfect phase control for both infinite and finite phase resolutions depends on the knowledge of exact CSI, which however is practically challenging to obtain due to the lack of signal processing capability at the IRS and the large size of passive scattering elements.

{\bf Summary}: In this part, we have reviewed the use of the IRS in wireless networks to minimize the transmit power or maximize the energy efficiency. The literature reveals an important power scaling law showing that the BS's power consumption can be scaled down in the order of $1/N^2$ without compromising the SNR at the receiver. The similar power scaling law still holds for a practical implementation of the IRS with low phase resolution. The power saving of IRS-assisted systems becomes more significant for wireless users far away from the transmitter. Though significant performance improvement can be verified by numerical results and simulations, we observe that the research focuses of almost all papers in literature are limited to the joint optimization of active and passive beamforming under different network scenarios. In fact, the overall performance gain can be better explored in the future work if the size of the IRS's scattering elements, the orientation of the IRS tiles, their partitions and grouping strategies, etc., are all taken into account, in combination with the transceivers' access control, user association, information encoding, transmit scheduling, QoS provisioning, etc.

More specifically, we notice that there are a few potential research problems that are not fully explored in the current literature and thus can be left for future study. Firstly, the power budget constraint of a practical IRS is barely mentioned in the literature. The EE/SE performance maximization relies on the characterization of total power consumption in IRS-assisted wireless networks. The power consumption of a practical IRS is not negligible and can be modeled by a linear function of its size and tuning resolution~\cite{19ee_chau}. However, most of the current researches consider self-sustainable IRS with sufficient energy supply, which may lead to over-optimistic conclusions. A preliminary work in~\cite{yuze} introduces the IRS's power budget constraint in a simple point-to-point MISO downlink system, which can be fulfilled by RF energy harvesting from the AP's signal beamforming. It is worthy further investigation in future research due to the importance of energy budget constraint in energy constrained IoT networks. Secondly, we may expect more research effort dedicated to performance maximization of MU systems under different multiple access schemes. The majority of current research works focus on interference-limited MU systems, where IRS is employed to assist spectrum sharing among different users and improve spectrum efficiency. The research of IRS-assisted multi-user access and performance maximization in different MAC protocols is very limited and should be explored more in the future. For example, we only observe one work in~\cite{ofdma} focusing on IRS-assisted RB allocation in an OFDMA system. Besides, we notice that IRS-assisted systems face the common challenges of channel estimation and the design approaches under incomplete or uncertain channel information. The authors in~\cite{two-timescale} show their effort along this research direction by designing a two-timescale sum-rate maximization algorithm that reduces the need for channel estimation. While a large majority of the existing works assume perfect channel information and focus on static maximization problems.

\subsection{IRS-assisted Physical Layer Security}

The IRS's wave manipulation has the flexibility of simultaneously creating enhanced beams to an intended receiver and suppressed beams to unintended receivers. This can be used to enhance physical layer security in wireless communications. The authors in~\cite{Yu2019Enabling} use IRS to defend against eavesdroppers. In particular, a single-antenna eavesdropper is located in the communication range between a multi-antenna legitimate transmitter (LT) and a single-antenna legitimate receiver (LR). To prevent the eavesdropper from eavesdropping by, the IRS placed near the LR can control the reflected signals to maximize the achievable secrecy rate at the LR, which is defined as the amount of information per time unit that can be securely sent over a communication channel~\cite{Oggier2011The}. The joint optimization of the LT's transmit beamforming and the IRS's phase shift matrix is approximately solved by BCD and MM algorithms, similar to that in~\cite{multigroup,18maxrate_chau}. Simulation results verify a significant improvement on the secrecy rate comparing to the cases without IRS. Besides, it is shown that it can be more efficient to enhance secrecy rate and energy efficiency by deploying large-scale IRSs instead of increasing the size of active antenna array at the transmitter. This idea has been extended in the recent research works by considering multiple antennas at the LR/eavesdropper~\cite{secrecy_max,secure-mimo}, rank-one/full-rank communication channels~\cite{secure_irs}, and secure THz communications~\cite{secure-thz}. Besides secrecy rate maximization, the authors in~\cite{chu-secure} minimize the transmit power subject to the secrecy rate constraint at the LR, by optimizing transmit power allocation and the IRS's phase shift matrix in the alternating optimization framework. The above research works assume implicitly that the eavesdropper is distant to the LR, or the eavesdropper has a worse channel condition, which is required to ensure the effectiveness of secure communications.

\begin{figure}[t]
	\centering \includegraphics[width=0.45\textwidth]{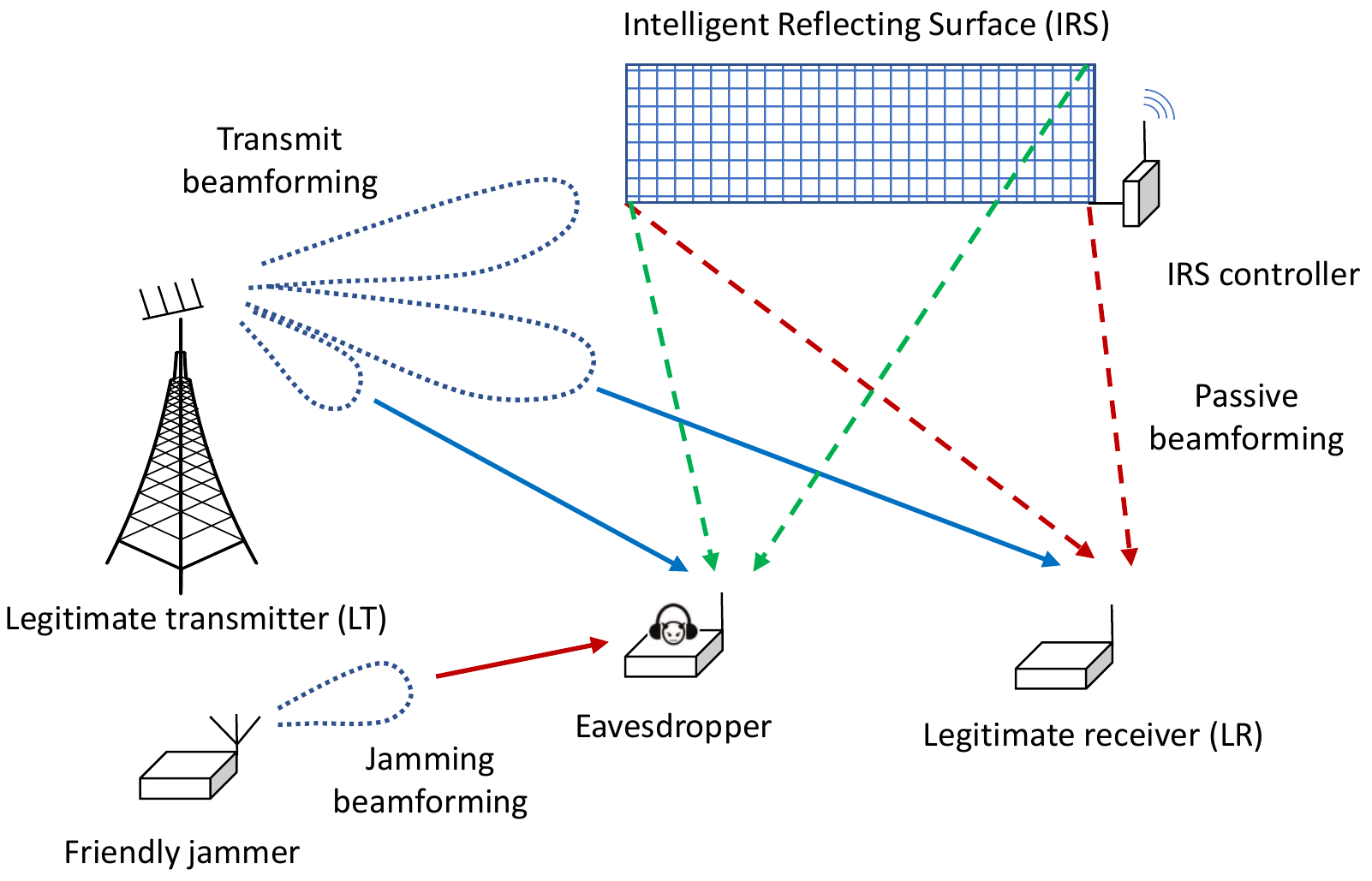}
	\caption{An IRS-assisted secure communication system with a friendly jammer.}
	\label{fig:Eav2}
\end{figure}

Considering a similar IRS-assisted secure wireless system to that in~\cite{Yu2019Enabling}, the authors in~\cite{secure_rui} focus on a more challenging scenario in which the eavesdropping channel is better than the legitimate channel and they are also highly correlated in space. This makes the achievable secrecy rate of the system very limited. To maximize the achievable secrecy rate, the optimal solution requires that the IRS's reflections and the LT's beamforming signals are destructively added at the eavesdropper. This idea is then extended to a more general scenario in which multiple eavesdroppers and LRs coexist in the system~\cite{19pls_liang}, where the authors maximize the users' minimum-secrecy-rate considering both continuous and discrete phase shifts at the IRS. This problem is approximately solved by the alternating optimization and path-following algorithms in an iterative manner. In practice, it is generally difficult to know the number, location, and channel conditions of eavesdroppers, and thus more practical solutions based on incomplete environmental information can be studied in future work. Another drawback of the above works is that the LT has very limited space to fight against the eavesdropper. It is obvious that the achievable secrecy rate is constrained by the network topology and channel conditions.

Besides the LT's transmit beamforming, the authors in~\cite{guan_secrecy} allow the LT to actively inject noise-like jamming signals to the channel, and further optimize the LT's jamming signals to enhance the achievable secrecy rate. By incorporating jamming signals into transmit beamforming, the secrecy rate is shown to be significantly improved compared to the conventional methods without using the IRS and/or jamming beamforming. The authors in~\cite{resource} consider a similar model and optimize the covariance matrix of artificial noise to deliberately impair the eavesdropper's channel, along with optimization of the BS's signal beamforming and the IRS's phase shift matrix based on SCA and SDR techniques. The numerical results show that the average secrecy rate can be improved by over 20\%, compared to the case without using dedicated artificial noise. The authors in~\cite{19pls_ian1} consider different approach to maximize the secrecy rate, by optimizing the number of IRS tiles (namely hypersurface tile) and their orientations to create desirable phase changes. In this way, the received signal power at the eavesdropper becomes very small even if the eavesdropper is located between the LT and LR. Though this is an effective solution to prevent eavesdropping, it requires the implementation of the hypersurface tile and thus its applications are limited to some particular scenarios such as smart offices and houses. All aforementioned works generally rely on the LT's optimal control (e.g.,~artificial noise injection and transmit beamforming) combined with the IRS's passive beamforming to improve the secrecy performance. Different from them, the authors in~\cite{wang2019eefficient} introduce an interesting idea by using a third device, namely a friendly jammer, to cooperate with the LT and to fight against multiple eavesdroppers, as shown in Fig.~\ref{fig:Eav2}. The secrecy rate maximization in this case requires a joint optimization of the LT's transmit beamforming, the friend's jamming beamforming, and the IRS's passive beamforming. The solution method follows a similar approach as those in~\cite{19pls_liang,Yu2019Enabling}, by leveraging the alternating optimization and SDR approaches.

The secrecy rate maximization merely aims at preventing legitimate transmissions from being deciphered or eavesdropped by an illegitimate user. However, this may not be enough for some cases where exposures of transmission activities, location and movement of transmitters are sensitive information to the end users. This calls for covert communications that can provide stronger protection by hiding the presence of legitimate transmissions~\cite{covert18}. The authors in~\cite{lu2019intelligent} propose the use of the IRS to enhance communication covertness, by leveraging IRS to reshape undesirable propagation conditions and thus avoid information leakage. A joint optimization of the LT's transmit power and the IRS's passive beamforming is proposed to maximize the achievable rate satisfying the covertness requirement. The numerical results demonstrate a significant increase in covert rate compared to the conventional cases without using the IRS. However, the efficacy of this work is based on the knowledge of perfect channel conditions.

{\bf Summary:} In this section, we have reviewed emerging applications of IRS for physical layer security. In particular, IRS can be used as a very effective tool to prevent wireless eavesdropping attacks by simultaneously controlling the transmissions at the LT and the reflection at the IRS. As a result, the achievable secrecy rates obtained by the IRS-assisted systems can be significantly improved compared with the conventional methods only relying on the LT's transmission control. Many simulation results verify the improvement of secrecy performance of IRS-assisted systems. This paves the way for a flourish of new research problems relating to physical layer security issues in the future wireless networks. However, there are still some challenges which need to be addressed. For example, how to simultaneously control the LT's transmissions and the IRS's reflections in real-time systems, and how to obtain the eavesdropper's channel information for accurate beamforming optimization are still major challenges for anti-eavesdropper systems.

\section{EMERGING USE CASES OF IRS IN WIRELESS NETWORKS}\label{sec_misc}

As demonstrated in previous sections, the use of the IRS is capable of bringing unprecedented performance enhancement for future wireless systems by reconfiguring the previously uncontrollable wireless channels in favor of network performance optimization. In particular, we have reviewed in previous sections the performance analysis and optimization of IRS-assisted wireless networks with different design objectives, i.e., SNR/capacity maximization, transmit power minimization, EE/SE performance maximization, and physical layer security issues. However, the potentials of using the IRS in wireless systems are not limited to the above aspects. It is still developing and flourishing in various aspects far beyond the aforementioned topics.

\subsection{Deep Learning for IRS-assisted Systems}

In Section~\ref{sec_optimization}, the joint active and passive beamforming design problems are mostly formulated as a non-convex problem and typically solved with sub-optimum in the alternating optimization framework. As the size of scattering elements becomes large, e.g.,~up to hundreds~\cite{irs_survey}, the optimization problem usually has high computational complexity and thus becomes difficult for practical implementation in a dynamic radio environment. Different from optimization methods, the authors in~\cite{19phase_chau} propose a deep learning (DL) method for efficient online reconfiguration of the IRS in a complex indoor environment. It is based on the fact that the IRS's optimal phase configuration depends on the receiver's location to maximize the received signal strength. To save time for online optimization, the deep neural network (DNN) is employed to construct a direct mapping between the receiver's location and the optimal phase configuration. The offline training of the DNN is based on a fingerprinting database recording the optimal phase configuration at each receiver's position. Similarly, the authors in~\cite{unsupervised} propose a DL approach for the IRS's passive beamforming optimization, which is based on a well-trained DNN to make real-time predictions. Simulation results show that the DL approach achieves close-to-optimal performance with significantly reduced time consumption compared to the conventional SDR-based methods. However, in the above learning based approaches, the optimal phase configuration in offline training is still achieved by exhaustive search or the alternating optimization method.

The authors in~\cite{drl-irs} propose a novel deep reinforcement learning (DRL) approach to address the secrecy rate maximization problem. DRL provides a mechanism to build knowledge from scratch and make autonomous decisions to improve network performance by continuously interacting with the environment~\cite{drl-hoang}. It can be very effective for complicated systems with diverse user requirements and time-varying channel conditions. However, a training process is still required to attain the decision intelligence. The DL approaches is also applied to the IRS's channel estimation problem, which typically incurs huge training overhead due to the large number of reflecting elements. The authors in~\cite{deep-mmwave} focus on channel estimation of an IRS-assisted massive MIMO system and propose a DL solution to learn how to optimally interact with the incident signal. The authors in~\cite{neural} construct an interpretable neural network for learning the optimal configuration of reconfigurable reflecting elements, which are modeled as the neural nodes in different layers of a back-propagating neural network. By a training period, the neural network can learn the propagation characteristics and adapt to facilitate the signal transmissions. Numerical evaluation shows that it can minimize the number of IRS tiles required for serving one transceiver pair. Practically, the training of DL approaches can be time consuming and unreliable, which may prevent its implementations in real systems.

\subsection{IRS-assisted Wireless Power Transfer}

The IRS's passive beamforming can be designed to enhance the received signal strength at an information receiver (IR). This approach can also improve the efficiency of wireless power transfer to an energy receiver (ER). The authors in~\cite{channel-wpt} consider a point-to-point MISO system for wireless energy transfer. The power beacon station's active beamforming and the IRS's passive beamforming are jointly optimized to maximize the signal power at the receiver. This incurs a trade-off between the size of active antenna array at the beacon station and the size of passive reflecting elements at the IRS. Considering multiple ERs, the authors in~\cite{powermax-swipt} maximize the weighted sum power at the ERs by jointly optimizing the BS's transmit beamforming and the IRS's phase shifts. A similar model is studied in~\cite{19miso_swipt}. The design objective is to maximize the ERs' minimum power subject to the IRs' SINR constraints and the AP's power constraint. A high-quality solution is obtained by SDR and alternating optimization.

The authors in~\cite{irs_swipt} consider an IRS-assisted simultaneous wireless information and power transfer (SWIPT) system, in which a multi-antenna BS communicates with several multi-antenna IRs, while guaranteeing the ERs' energy harvesting requirements. The authors firstly formulate a weighted sum-rate maximization problem by jointly optimizing the BS's transmit precoding and the IRS's phase shifts. The classic BCD algorithm is adopted to find a near-optimal solution, which is much better than those of baselines, i.e., with fixed phase or without using the IRS. A special case of the IRS-assisted SWIPT is studied in~\cite{cooperation}, where two users firstly harvest energy and then transmit information to an AP following the TDMA protocol. To maximize the minimum throughput, the authors in~\cite{cooperation} propose a joint optimization of the users' time and power allocations, as well as the IRS's passive beamforming in wireless energy and information transfer. The numerical results reveal that on average the IRS-assisted case can achieve over 74\% throughput increase compared to the case without IRS. The authors in~\cite{lyubin} consider a more general MU MISO scenario, where the IRS firstly assists downlink energy transfer to the users and then enhances the users' uplink information transmissions in the TDMA protocol. A sum-rate maximization problem is formulated by jointly optimizing the user's transmission scheduling, the IRS's phase shift matrices for energy and information transfer. Simulation results verify that the proposed scheme can improve the sum-rate by 350\% compared to the case without IRS. The authors in~\cite{wu2019joint} use a set of distributed IRSs to assist SWIPT from a multi-antenna AP to multiple IRs and ERs. Instead of sum-rate maximization, the authors in~\cite{wu2019joint} focus on the AP's transmit power minimization by jointly optimizing the AP's transmit beamforming and the IRSs' passive beamforming, subject to the IRs' SINR constraints and the ERs' energy constraints. Simulation results demonstrate the significant performance gains achieved over benchmark schemes, e.g., the AP's transmit power can be reduced by more than 50\% with only 30 scattering elements in the demonstrated setup.

\subsection{IRS-assisted UAV Communications}

The capacity maximization problems in~\cite{capacity_rui} and~\cite{irs_ofdm} can be extended to the emerging UAV networks. In~\cite{uav_bf_marco}, the authors introduce an IRS-assisted UAV communication network in which the on-building IRS is used to enhance communication quality from UAV to the ground user. A joint optimization of the UAV trajectory and IRS's passive beamforming is formulated to maximize the average achievable rate. Given the UAV trajectory, the IRS's phase shift is firstly derived in a closed form. Then, with the fixed phase shifts, the local-optimal trajectory solution can be derived by using the SCA method. The authors in~\cite{ma2019enhancing} employ the wall-mounted IRS to enhance the channel between cellular BS and UAVs, which previously suffers from poor signal strength as the BS's signal beamforming is generally optimized to serve ground users. By controlling the IRS's reflecting phase, the signal gain at the UAV is characterized based on the 3GPP ground-to-air channel models as a function of various deploying parameters, including the UAV's height, the IRS's size, altitude, and distance to the BS. Thus, the maximum signal gain can be achieved by optimizing the IRS's location, altitude, and distance to BS. Numerical results show that a significant signal gain can be achieved for UAVs even with a small-size IRS, e.g.,~the signal gain quickly jumps to 20 dB when UAVs fly over 30 meters above the BS.

The authors in~\cite{aerial} deploy aerial IRS to enjoy full-angle reflection and LOS channel conditions. The aerial IRS can be carried by balloons or UAVs and used to enhance signal coverage of a cellular network. The maximization of the worst-case SNR in a coverage area on the ground is formulated as an optimization of the transmit beamforming, the placement, and phase shift matrix of the aerial IRS. A similar scenario is studied in~\cite{uav_carry_irs} where the UAV-carried IRS is used to enhance the transmission performance of mmWave communication networks. The UAV-carried IRS is also capable of energy harvesting from the mmWave signals to sustain its operations. A reinforcement learning (RL) approach is proposed to find the optimal policy, i.e., the best location of UAV, that maximizes the average throughput of the downlink transmissions. The simulation results show that the RL-based approach can improve the network performance by 65\% compared to conventional schemes without learning capability.

\subsection{IRS-assisted Mobile Edge Computing}

MEC allows data and computation offloading to resource-rich MEC servers. As such, the energy consumption and processing delay can be potentially reduced at the end users with insufficient computing capability. However, the benefit of MEC is not fully exploited, especially when the link for data and computation offloading is hampered. As the use of IRS can enhance both the EE/SE performance, it can be a promising technology to improve the MEC performance.

The authors in~\cite{computation} propose an IRS-assisted MEC system where multiple users report individuals' data to the AP for data aggregation. The offloading process is assisted by the IRS's passive beamforming to enhance the channel conditions. The design target is to minimize the worst-case mean-squared-error in data aggregation, which is solved by a novel alternating difference-of-convex (DC) programming algorithm. The authors in~\cite{hua2019reconfigurable} use the IRS in a green edge inference system, where the computation tasks at resource-limited user devices can be offloaded to multiple resource-rich BSs. To minimize the network power consumption in both computation and uplink/downlink data transmissions, the authors in~\cite{hua2019reconfigurable} propose a joint optimization of the task allocation among different BSs, each BS's transmit and receive beamforming vectors, the users' transmit power, and the IRS's passive beamforming strategies. The authors firstly propose a reformulation by exploiting the group sparsity structure of the beamforming vectors, and then decouple the optimization variables by a block-structured optimization approach. Instead of the widely used SDR approach, a novel DC-function based three-stage framework is introduced to solve the original problem with enhanced network performance. Numerical results reveal that the proposed approach can reduce the overall power consumption by around 20\% compared to the conventional SDR-based approach.

{\bf Summary}: The IRS is a cutting-edge technology possessing outstanding features expected to open new promising research directions, which have never been seen before in wireless communication networks. In this section, we have reviewed some emerging applications of the IRS in wireless networks including wireless power transfer, UAV communications, and MEC. It can be clearly seen that by using the IRS, energy, communications and computing efficiency of conventional wireless networks can be significantly enhanced. However, all results obtained so far are based on simplified models and verified through simulations. More proof-of-concept prototypes are required to validate the IRS's practical efficiency. Besides, there are more potential applications that can be enhanced by using the IRS, which can be studied further in the future work, such as vehicular and maritime communications, satellite and next-generation mobile networks.

\section{CHALLENGES AND FUTURE RESEARCH DIRECTIONS}\label{sec_open}
Different approaches reviewed in this survey evidently show that IRS-assisted wireless networks can effectively enhance the received signal power, extend the network coverage, increase the link capacity, minimize the transmit power, suppress the interference, and enable better security and QoS provisioning to multiple users, etc., compared to non-IRS-assisted counterparts. However, there still exists some challenges, open issues, and new research directions which are discussed as follows.

\subsection{Challenges and Open Issues}

\subsubsection{Energy-efficient Channel Sensing and Estimation}
The IRS is composed of a large array of passive scattering elements which are typically interconnected and controlled by a centralized controller, e.g.,~\cite{16surface_focusing,tuning18,18surface_circuit}. The superiority of using the IRS relies on its reconfiguration of each scattering element's phase shift, according to the channel conditions from the transmitter to its receiver. This requires the capability of channel sensing and signal processing, which becomes very challenging without dedicated signal processing capability at the passive scattering elements. The channel estimation of an IRS-assisted system is typically performed at one end point of the communication process, e.g.,~the BS with higher computation capability. Existing approaches for the IRS's channel estimation generally assume that only one scattering element is active each time, while all the other elements are inactive, e.g.,~\cite{channel-wpt,channel-backscatter}. Such an element-by-element ON/OFF-based channel estimation scheme is practically costly for a large-scale IRS with massive scattering elements. In particular, the IRS is not fully utilized as only a small portion of the scattering elements is active in each time. This degrades the channel estimation accuracy and produces a long estimation delay. The authors in~\cite{chestimate} propose the optimal activation pattern for the IRS's channel estimation, which achieves one order lower estimation variance compared to the ON/OFF-based methods. The authors in~\cite{wang2019channel} and~\cite{ofdm_estimation} assume that all reflecting elements are switched on. By setting specially designed phase shift at each reflecting element, the IRS-enhanced uplink channel can be estimated efficiently with the reduced length of pilot sequence. More sophisticated signal processing algorithms are also designed for channel estimation, e.g.,~\cite{xia2019intelligent,ofdma-channel,cascade}, aiming at achieving better accuracy or lower training overhead, however they generally demand higher power consumption for information exchange, signal processing, and computation. A practically efficient and sustainable channel estimation is still one of the key enabling technologies for IRS-assisted wireless systems.

\subsubsection{Practical Protocols for Information Exchange}

Generally, the IRS's channel sensing and estimation can be achieved by overhearing a training sequence sent by the active transceiver. Thus, information exchange between IRS and the active transceiver is required to synchronize the overhearing and pilot training. Information exchange also happens when a transmission scheduling protocol is employed to coordinate MU's data transmissions. In this case, the IRS also needs to synchronize with different transmission frames and reconfigure its passive beamforming schemes according to different users' channel conditions. A practical protocol is thus required for the IRS to talk with the conventional transceivers. Information exchange can be made easy for conventional transceivers using a dedicated control channel. However, without sufficient energy supply, it becomes more challenging for the passive IRS to detect and decode the information from other active transceivers. Hence, the design of an information exchange protocol firstly has to be of extremely low power consumption such that it is sustainable by wireless energy harvesting. Secondly, it has to be cost-effective by minimizing the conflict with or the alteration of the existing systems. A potential design idea can be incurred by the IRS's inherent sensing capability~\cite{19survey_renzo}. In particular, it can be more energy-efficient for the IRS to sense physical layer information, instead of decoding MAC layer information bit streams. Therefore, the information exchange can be made possible by modulating the packet length or transmit power, so that the passive IRS can sense the variations of signal power with low-power consumption.

\subsubsection{Reflection as a Resource for IRS-assisted HetNets}

In the future smart radio environment, the wireless networks can be assisted by a distributed IRS system with individually controlled IRS units due to the pervasive deployment of reconfigurable metasurfaces on different objects. This implies a challenging situation for the real-time allocation and optimization of different IRSs to serve multiple data streams in dynamic and heterogeneous networks (HetNets). Conventionally, individual transceivers can independently adapt their operational parameters to the channel condition, which follows some stochastic model and can be predicted or estimated via a training process. However, with the IRS's reconfigurability, the radio environment itself becomes controllable and non-stationary. Hence, it becomes more difficult for individual transceivers to understand the CSI via training. This implies a centralized coordination for the IRS-assisted networks, at least for the distributed IRS units. This makes the wireless environment tractable and controllable. Specifically, a joint control mechanism is required for efficient allocation and association of IRS units to serve multiple users simultaneously.

\subsubsection{Agile and Light-weight Phase Reconfiguration}
The phase control of an individual scattering element has to be coordinated with each other for effective beam steering. The large size of the IRS's scattering elements can make the overall phase tuning more flexible, even with limited phase shifts at an individual scattering element, e.g.,~\cite{stat-csi,discrete-shift,guo2019weighted,19pbf_rui2}. Such a flexibility comes with a cost. In particular, an efficient algorithm is required to jointly control the phase shifts of all scattering elements in a timely manner, according to the dynamics of the radio environment. The increase in the size of scattering elements also imposes great pressure to channel estimation, making it even harder for efficient phase control. Besides, the IRS's phase control is strongly coupled with the transmit control of the active transceivers. This makes it more challenging for the design of an agile and light-weight phase control algorithm with minimum energy consumption and communication overhead. In the current literature, most of the phase control algorithms are based on alternating optimization method, that decouples the IRS's phase control and the conventional transmit control (e.g., power allocation, transmit beamforming, and precoding matrix design) in separated sub-problems. Though this simplification can provide a convergent solution to a sub-optimum, it inevitably incurs large communication overhead and processing delay. It is also very challenging to characterize the performance loss of the convergent solution compared with the optimum.

\subsection{Future Research Directions}

Based on extensive literature review and the analysis of common shortcomings of the current literature, we highlight a few potential research directions for future exploration.

\subsubsection{Learning Approach for Passive Beamforming}Different from the alternating optimization commonly used in the literature, machine learning approaches can be more appealing for the IRS to realize agile and light-weight phase control based on locally observed information of the radio environment~\cite{19learning_renzo,unsupervised,model_dnn19}. This can help minimize the overhead of information exchange between the IRS and active transceivers. The large number of scattering elements and their sensing capabilities further imply that rich information can be collected during channel sensing, providing the possibility for data-driven DL approaches~\cite{19survey_renzo,dlsurvey,19phase_chau,unsupervised}. Furthermore, the potential analog computation can be also envisioned to realize ANNs via multi-layer metasurfaces~\cite{14science,neural}, which potentially make the learning approach agile in computation and light-weight without the need for information exchange. However, current DL approaches are still facing many practical challenges, including the training overhead, stability and adaptability issues. The design of DL approaches has to meet the hardware constraints of IRS-assisted wireless systems, such as limited computation and communication capabilities of the passive scattering elements. For example, leveraging RL approaches, a decision-making agent can be employed at the IRS controller to adapt its phase configuration, solely based on the observed system state (e.g.,~the perceived CSI via its sensing capability) and the receivers' feedback of its phase configuration, e.g.,~\cite{drl-irs,uav_carry_irs}. The system state can be estimated by the IRS via sensing or overhearing the ACK packets from the receiver to the transmitter. With specially designed ACK packets, e.g.,~the ACK packets with different time durations or transmit power, the channel sensing of IRS can be made easier without energy consumption on decoding the ACK packets. A similar idea has been used for information exchange in wireless backscatter communications~\cite{ambient_survey}.

\subsubsection{IRS-assisted D2D Communications}
D2D communications technology is envisioned to connect billions of low-power user devices. Different from the typically downlink transmissions from multi-antenna AP to receivers, D2D communications become more decentralized and diverse, which brings new research problems for IRS-assisted D2D communications. In one aspect, The IRS can be dynamically reconfigured to enhance individual data link of D2D communications. This requires highly efficient channel sensing and estimation protocols, as well as agile phase reconfiguration algorithms. The insufficiency of energy supply for the IoT devices implies another difficult situation that demands minimized interactions between the IRS and the IoT devices. In another aspect, the distributed IRS units can be used to understand the system profile by learning from a large amount of IRS-assisted transmissions in a spatial-temporal region~\cite{mobilebigdata}. The system profile may include the information about the potential bottleneck devices, the time-varying traffic pattern, the energy distribution over the entire network, and the information for predicting and diagnosing network failures. Such information can be further used by the D2D networks to optimize the deployment and settings of the IRS units, the IoT devices' transmission control, the placement of relay nodes, and the power beacon stations.

\subsubsection{IRS-assisted mmWave and THz Communications} One of the promising applications of the IRS is in the extended coverage of 5G and beyond 5G communications. We expect that mmWave 5G communication and future THz beyond 5G communication will face with a critical issue of dead-spots which will not be covered well because of the severe blocking loss of such short-length waveforms. In such situations, the IRS's two salient EM properties of reflection and refraction can be exploited to resolve the critical issue of dead-spots when IRSs are deployed in between the base stations and end users. For example, a user is located in the same side of the serving BS, in which case the incident EM wave on the IRS can be reflected toward the user, whereas if a user is in the opposite side, then the incident EM wave can be refracted through the IRS to reach the user with enhanced signal quality. It is envisioned that the 3D deployment of IRSs with 5G and beyond 5G wireless systems for eliminating such dead-spots will be cost-effective, and the current massive MIMO and mmWave technologies evolving will be integrated with the IRS technology for the extended coverage eventually.

\subsubsection{Using IRS in Smart Wireless Sensing} The current research typically uses the IRS as an auxiliary way for enhancing transmission performance of the existing transceivers. In fact, each scattering element of the IRS can be individually phase-tuned and thus showing different sensitivities to the incident signals from different directions. This implies that the IRS can be employed as an array of sensor devices that are configured to passively monitor the radio environment~\cite{19survey_renzo}. Given wired or wireless connections to a centralized IRS controller, all the sensing information from different scattering elements can be collected and analyzed jointly in an energy-efficient way. From this view point, the use of the IRS as an array of smart sensors will have rich applications in wireless sensing, e.g., indoor positioning~\cite{18positioning_sha,he2019adaptive} and human pose understanding~\cite{rf_pose}. As such, IRS-assisted wireless systems will not only enhance communications but also bring the possibility of human-network interactions, i.e., the communication performance and user satisfaction can be even better by using IRS to understand the human behavior or intention in wireless networks~\cite{bigdata_human_in_loop}.

\subsubsection{Trade-off between Array Gain and LOS Path Loss} Using the IRS as passive relay provides the array gain $N^{2}$ thanks to no noise addition when $N$ reflecting elements are employed. In addition, if the reflecting element is of sufficiently large size (e.g., $10\lambda\times10\lambda$ for the wavelength $\lambda$), the IRS can act as the specular reflector like lens in which case the path loss follows the ``sum-distance" path loss, unlike the active relay whose antenna size is on the order of $\lambda$, resulting in the severe ``product-distance" path loss. Therefore, there exists a crucial trade-off between achieving a larger array gain and guaranteeing the minimal LOS path loss because the number of reflecting elements varies per unit area depending on the size of the reflecting elements. Namely, to assure the LOS path loss, we have to sacrifice the array gain whereas the largest array gain can be achieved while compromising the LOS path loss. The latter is due to the minimum physical size of a reflecting element (like lens) that focuses the energy onto a focal point depending on the distance between the IRS and the receiver (i.e., focal point). Future research for characterizing the trade-off, considering the 3D deployment of IRSs, will be of paramount importance, in that the smart radio environment can be fully utilized in terms of the density of IRSs and their sum gain, normalized by the implementation cost.

\subsubsection{Environment AI for Smart Wireless} IRS can be used as one of the following three functions: 1) Passive Relay, 2) Passive Transmitter, 3) both of them, where the quality of primary signal is enhanced by passive beamforming via relay and at the same time, the secondary information generated from the IRS itself can be embedded in the primary signal (like ambient backscatter), e.g.,~\cite{kim,channel-backscatter,pbit19,sharing}. For example, the IRS may be equipped with sensors monitoring environments, which generate such secondary information to be reported to the IoT gateway in the uplink. Therefore, the mode switching at IRS will need to be intelligently and remotely performed by the control center through the IoT gateway, e.g.,~edge node, considering the user objectives and device positions. Moreover, if a large number of IRS is deployed in a certain area to assist the primary transmission while transmitting their own secondary information (from IoT sensors), the global control of these IRSs will need to be handled by the control center by gathering the user objectives and device positions, so as to assure the optimal routing of air routes of IRSs in conjunction with the mode switching. However, due to the latency and privacy issues, the global control by the control center may not be feasible. Instead, the collaborative (federated) learning will play a crucial role to intelligently perform the required global control through the cooperation with the edge nodes, which perform the learning locally and upload their model parameters to the control center. This way we can resolve the latency and privacy issues.

\section{CONCLUSIONS}\label{sec_cons}
\label{sec:conclusion}

This paper has presented a comprehensive survey on the design and applications of the IRS to wireless communication networks. Firstly, we have presented an overview of the metasurface and its reconfigurability to realize the vision of IRS. Then, we have focused on its applications in wireless networks and reviewed different network scenarios that can benefit from its reconfigurability. Afterwards, we have provided detailed reviews on the performance analysis and optimization of IRS-assisted wireless networks under different communication scenarios, including SNR/capacity maximization, transmit power minimization, EE/SE performance maximization, and secrecy rate maximization, etc. Due to its diversified applications, we have also reviewed the emerging use of the IRS to promote wireless power transfer, UAV communications, and MEC. Finally, we have outlined important challenges, open issues as well as future research directions.

This survey covers a broad set of research topics regarding IRS and its applications in wireless networks, from its physical characterization and channel modeling, to research problems in a wireless networking perspective, mainly focusing on the optimization models and solution methods for IRS-assisted wireless systems. The current literature has identified basic problem models and solution methods for IRS-assisted wireless systems, including tentative discussions on practical limitations, such as the IRS's finite phase resolution, partial channel information, and hardware impairments. In the current literature, a large majority of the solution methods are based on alternating optimization framework, which ensures the convergence to a local-optimal solution. Though significant performance improvement can be verified by numerical results and simulations, the performance gap with the optimum is not known exactly and seldom characterized in literature. In the future work, by exploring more sophisticated optimization algorithms, the IRS-assisted wireless systems are expected to achieve a higher performance gain than that in the current literature. Besides, the current research works are mainly limited to joint optimization problems of active and passive beamforming under different network scenarios. The overall performance gain of IRS-assisted wireless systems can be better explored in the future work if the size and distribution of the IRS's scattering elements, the orientation and mobility of the IRS tiles, their partitions and grouping strategies, etc., are all taken into account, in combination with multi-users' access control, user association, information encoding, transmit scheduling, QoS provisioning, etc. This will open a significantly larger space to explore than the research scope in the current literature.

\bibliographystyle{IEEEtran}
\bibliography{irsurvey}{}


\end{document}